\documentclass[twocolumn]{aastex631}

\let\tablenum\relax

\usepackage{natbib,siunitx,url,graphicx,amsmath,footnote,float,xcolor,cases,lipsum}
\usepackage{subcaption}

\begin{document}
\let\WriteBookmarks\relax
\def\floatpagepagefraction{1}
\def\textpagefraction{.001}

Accepted for publication in Icarus

\title{The fragility of the Uranian moons during the giant planet instability}

\author[0000-0001-8933-6878]{Matthew S. Clement}
\affiliation{Johns Hopkins APL, 11100 Johns Hopkins Rd, Laurel, 20723, MD, USA}  

\author[0000-0001-5272-5888]{Nathan A. Kaib}
\affiliation{Planetary Science Institute, 1700 E. Fort Lowell, Suite 106, Tucson, AZ 85719, USA}
\affiliation{HL Dodge Department of Physics $\&$ Astronomy, University of Oklahoma, Norman, OK 73019, USA}

\author[0000-0003-1878-0634]{André Izidoro}
\affiliation{Department of Earth, Environmental and Planetary Sciences, 6100 MS 126, Rice University, Houston, TX 77005, USA}

\author[0000-0001-6730-7857]{Rogerio Deienno}
\affiliation{Solar System Science $\&$ Exploration Division, Southwest Research Institute, 1301 Walnut Street, Suite 400, Boulder, CO 80302, USA}

\correspondingauthor{Matt Clement}
\email{matt.clement@jhuapl.edu}

\begin{abstract}

It is thought that, some time after their formation on more compact orbits, the solar system's giant planets experienced a dynamical instability that caused their orbits to excite, diverge, and ejected one or more objects with masses comparable to the modern ice giants.  A key feature of this model is that the planets experience encounters with other planetary bodies, and these encounters facilitate the capture of nearby small bodies as irregular satellites.  Instability simulations indicate that planet-planet encounter distances can typically fall below 0.1 au, which is only roughly an order of magnitude larger than the radial extent of the modern planets' regular satellite systems.  In this paper we model the effects of these encounters on the dynamical stability of the regular moons of Jupiter and Uranus.  In contrast to previous studies that were limited to a narrow range of possible evolutionary tracks, we tested encounter histories from 122 plausible outer solar system dynamical histories.  We find that the survival probability for the Jovian and Uranian moon systems are both less than 15\%.  Moreover, we only identify one case where both Uranus and Jupiter’s large satellites consistently survive the same instability.  Interestingly, Jupiter's moons are most likely to survive in instabilities initialized with two smaller extra ice giants, and cases with one larger additional planet provide more favorable conditions for Uranian system survival.  In either case, if Uranus encounters another ice giant at $D_{enc}<$0.02 au, or one of the gas giants at $D_{enc}<$0.1 au, satellite system destruction is effectively guaranteed.  Wider encounters can also affect the system, particularly when they occur successively.  Since the Laplace resonance likely would not be in place today if Jupiter's moons were destabilized and collided with one another after their formation, our results indicate that Uranus' moons were likely perturbed to the point of collisions at least twice: as a result of both the impact that tilted the planet and the giant planet instability.  In the latter event, we find that neighboring moons likely underwent a series of high velocity hit-and-run collisions that redistributed material until the satellite's orbits were tidally recirculated.  We hypothesize that such a series of collisions might provide an interesting explanation for Miranda's small size and icy composition.  Alternatively, it is also possible that Uranus fortuitously avoided deep planetary encounters in a manner not captured in our sample of simulations.

\end{abstract}

\section{Introduction}
\label{sect:intro}

Jupiter and Saturn must have formed within just a few million years after the Sun's birth in order to accrete their massive gaseous envelopes before nebular gas dissipation  \citep{pollack96,haisch01}.  Given their similar properties (i.e. total mass, separation from the planet, equatorial orbits, etc) the large regular moons of both gas giants are also expected have accreted within a circumplanetary disk during this time \citep{lunine82,canup02,canup06,batygin_morby20,madeira21}; effectively as a consequence of formation of the planet itself.  However, it is not clear if Uranus and Neptune's formation can be neatly reconciled within this paradigm.  On the one hand, numerical models of giant planet formation via direct gravitational collapse \citep{boss97,mayer02} or pebble accretion \citep{lambrechts14a,bitch15,levison15_gp,ndugu18} can produce smaller cores that might struggle to efficiently accrete gas from the nebula and eventually yield good analogs of the ice giants \citep[e.g.][]{lambrechts17,raorane24}.  On the other hand, their $\sim$several-Earth mass heavy element cores \citep[as inferred from internal models:][]{nettelmann13} and large obliquities \citep{slattery92,boue10,rogosziniski20} have been interpreted as evidence of their having undergone a more prolonged epoch of growth involving one or more giant impacts with other large proto-planets \citep{jakubik12,izidoro15_ig}.  Such impacts have the potential to reshuffle or destroy any primordial satellite system that might have formed around the ice giants \citep{morby12_uranus,gomes24}, and potentially trigger the formation of a second generation of moons \citep{chau21}.

Sometime after the dissipation of the Sun's primordial gas disk, the giant planets' orbits and satellite systems were reshaped by the solar system's giant planet instability, also referred to as the Nice Model \citep{tsiganis05,gomes05,morby05}.  The instability's global consequences, and its ability to reproduce many specific qualities of the solar system's small body populations \citep{nesvorny13,nesvorny15a,nesvorny15b,nesvorny16,kaib16,nesvorny18,deienno18,deienno24}, the orbits of the planets themselves \citep{nesvorny12,roig16,deienno17,clement21_instb,clement23_merc5} and other geophysical constraints \citep{zellner17,morby18,brasser20,nesvorny23,edwards24} in numerical simulations are well documented.  

Of particular relevance to our current investigation, the capture of irregular satellites at all four giant planets is a natural consequence of the scenario \citep{nesvory07}.  Indeed, the Nice Model's capture mechanism largely supplanted previous proposed models for capture \citep[][each of which had their own documented issues]{columbo71,heppenheimer77,pollack79,philpott10} and remains one of the most convincing aspects responsible for the model's popularity \citep[see][for a recent review]{morby20_review}.  As the giant planets orbits evolve chaotically during the instability they experience close encounters with one another.  These epochs of close approach produce unique three body geometries that alter the trajectories of scattered planetesimals in the region (typically those already on hyperbolic orbits) in a manner that allows for capture \citep{nesvorny14a}.  While these occasional encounters are vital for capturing the irregulars with a reasonable degree of efficiency, \citet{deienno14} showed that they can potentially destabilize, and even destroy the regular moons.

Several peculiar qualities of the satellite system's of Saturn, Uranus and Neptune have been interpreted as strong evidence of late reorganization (i.e. after gas dissipation).  Perhaps most famously, it has been proposed that Saturn's current complex and intertwined system of rings and moons is much younger than the solar system \citep{cuzziandestrada98,tajeddine17,iess19,cuk26}. Indeed, \citet{cuk16} went as far as to propose that the low tidal Q of Saturn inferred from historical observations of tidal recession and high heat flow detected on Enceladus by Cassini \citep{porco06,lainey12} implies that Tethys, Dione, Rhea, Mimas, Enceladus and the rings all formed in one or more dramatic events within the last 100 Myr \citep[see also][]{lainey20}.  However, such a scenario is not consistent with the inferred ancient ages of these moon's surfaces from crater production models \citep{zahnle98,zahnle03,wong19,wong21,wong23,bottke24}.  The lack of large regular moons at Neptune also implies a curious origin story.  If Triton was captured from an encounter with a Pluto-Charon-like binary \citep{agnor06} or as a consequence of planetary encounters during the Nice Model \citep{nesvorny07}, it would likely have catastrophically destroyed any primordial regular satellites.  However, it is also possible that Triton and Nereid \citep{brown98_nereid} are the lone survivors of an embryo impact \citep{gomes24} that endowed the planet with its 30$^{\circ}$ obliquity.

As the 4:2:1 Laplace resonance between Io, Europa and Ganymede is a natural outcome of many satellite formation models, the modern Jovian system and resonance are typically assumed to resemble their post-formation state.  Early models for the formation of the resonance favored it assembling within the last few Gyr (depending on the value of Jupiter's tidal quality factor) via outward convergent tidal migration of Io and Europa followed by resonant capture, and subsequent migration of the resonant pair into resonance with Ganymede \citep{yoder79,greenberg87,malhotra91}.  In contrast, modern models \citep[e.g.][]{canup02,shibaike19,batygin_morby20,madeira21} favor a primordial origin via outside-in gas driven migration while the moons are forming within the circum-Jovian disk.  In particular, the substantial water-ice contents of the inner moons \citep[e.g.][]{kuskov01} are difficult to reconcile within a scenario where the moons largely form in-situ before tidally migrating into resonance.  Additionally, isotopic ratios in Io's atmosphere are consistent with substantial volcanic out-gassing driven by intense tidal heating over the age of the solar system \citep{dekleer24}.  This implies that the Laplace resonance has been in place since the solar system's birth.  However, the reason for Callisto's absence from the chain is still debated.  The simplest explanation for this would be that Callisto either formed too late or did not migrate fast enough to attain a resonant orbit before the circumplanetary disk dissipated \citep{peale02}, though tidal migration out of a primordial resonance \citep{fuller16,shibaike19,madeira21}, disruption during the giant planet instability \citep{deienno14} and sub-structures within the disk \citep{yap25} have also been proposed.  

The equatorially prograde nature of Uranus' regular moons seems to imply a connection between their formation and the event responsible for tilting the planet.  A number of tilting-mechanisms such as spin-orbit resonances with existing \citep{boue10,rogosziniski20} or hypothetical \citep{lu22} planets, the outward migration of a primordial satellite \citep{saillenfest22}, and impacts \citep[e.g.:][]{slattery92} have been proposed.  The most durable model has proved to be that of \citet{morby12_uranus}, where Uranus is tilted in a series of multiple giant impacts. The last of these obliquity shifts generates a debris field from a disrupted system of primordial satellites; within which the modern moons subsequently coalesce.  It has been argued that this impact must have occurred \textit{after Nebular disk dissipation}; otherwise an additional generation of satellites would have formed in Uranus' orbital plane via disk infall, destroying the equatorial satellites in the process \citep{rufu22}.  However, it remains uncertain whether the ice giants had the ability to form a circumplanetary disk via infall \citep{szulagyi18}.  Critical to this scenario is the formation of two separate disks (nominally one from the disrupted primordial satellites and one from the impact ejecta) such that the inner disk artificially enhances Uranus' J2 and realigns the outer disk with the planet's equator.  However, recent work taking in to account the planet's rotation concluded that the typical impacts invoked in the canonical \citet{morby12_uranus} scenario violate constraints on the final system angular momentum by a factor of 2-4 \citep{chau21,rufu22}.  This could be resolved if the primordial Uranian satellite system was smaller and more radially compact, or if the impactor's mass was lower; thus necessitating a lower mass inner disk post-impact.  Such a low-mass impact-produced disk has also been proposed as a potential explanation for Miranda's icy composition \citep[$\sim$23\% rock as opposed to $\sim$50\% for the other satellites][]{hussmann06,salmon22}.  However, it remains to be demonstrated how a compact system could have spread so substantially post-impact.

In this paper we investigate the possibility of an additional disruption of the primordial regular moons of Uranus transpiring during the epoch of planetary encounters.  The system's survival was studied in \citet{deienno11} in the context of what is now an outdated version of the Nice Model \citep[see][for a detailed explanation of the evolution of the different iterations of the model]{nesvorny12}.  The authors concluded that the satellites are particularly sensitive to encounters with Jupiter that occurred in some realizations, and that planetesimal flybys within the orbit of Oberon could also destabilize the system.  Similarly, \citet{deienno14} modeled the stability of the Galilean moons using three different instabilities of the same type as those that remain favored today \citep[e.g.][]{nesvorny21_tp}.  They found one of the three cases was successful in terms of retaining all four moons and not over-exciting their orbits.  Moreover, the authors found that encounters with ice giants within 0.02 au, or chains of repeated encounters $<$ 0.03 au, rapidly destroy the system. 

In contrast to these previous studies that were limited to only a handful of potential instability encounter sequences, here we consider a sample of 122 potentially viable outer solar system evolutionary tracks from \citet{clement21_instb,clement21_instb2}; each of which produce adequate analogs of the modern outer solar system.  We find that planetary encounters at Uranus are systematically stronger and more frequent than at the other three giant planets.  By studying the effects of these encounters on the regular moons orbits with a large suite of N-body simulations, we will argue in the subsequent sections that it is highly unlikely that the Uranian moons were not destabilized during the giant planet instability.  While we contrast these results with a smaller set of simulations of Jovian satellite survival, we choose to relegate the complete discussion of the instability's effects at Jupiter and Saturn, as well as the potential outcome of an instability among Uranus' primordial moons to a forthcoming series of companion manuscripts.

\section{Methods}

\subsection{Analysis of sample of instability evolutionary tracks}
\label{sect:methods1}

\begin{figure*}
    \centering
    \includegraphics[width=0.49\linewidth]{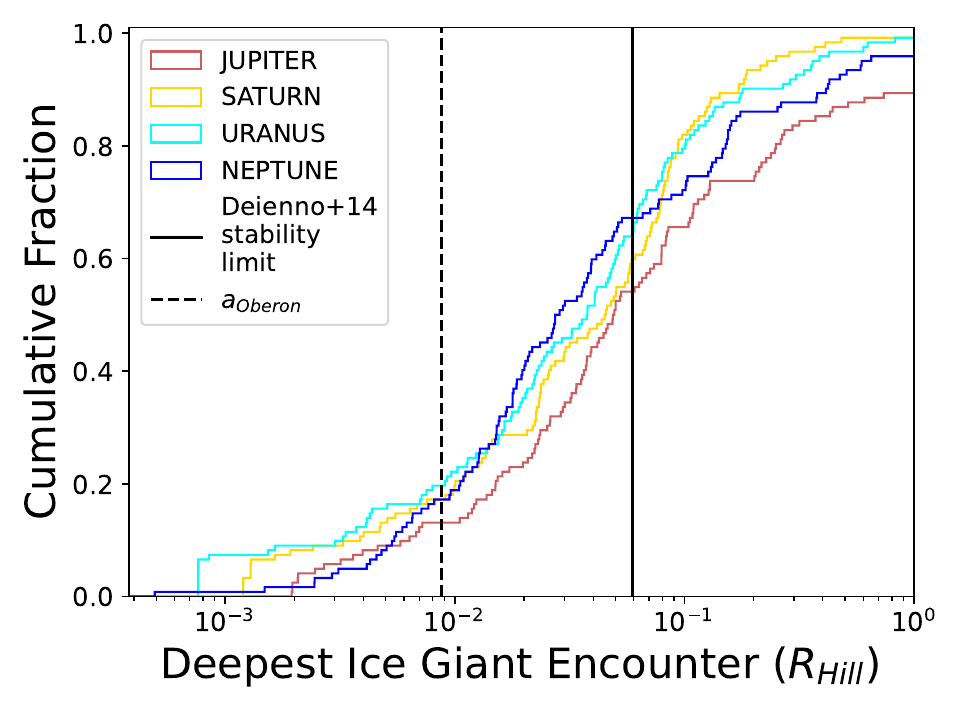}
    \includegraphics[width=0.49\linewidth]{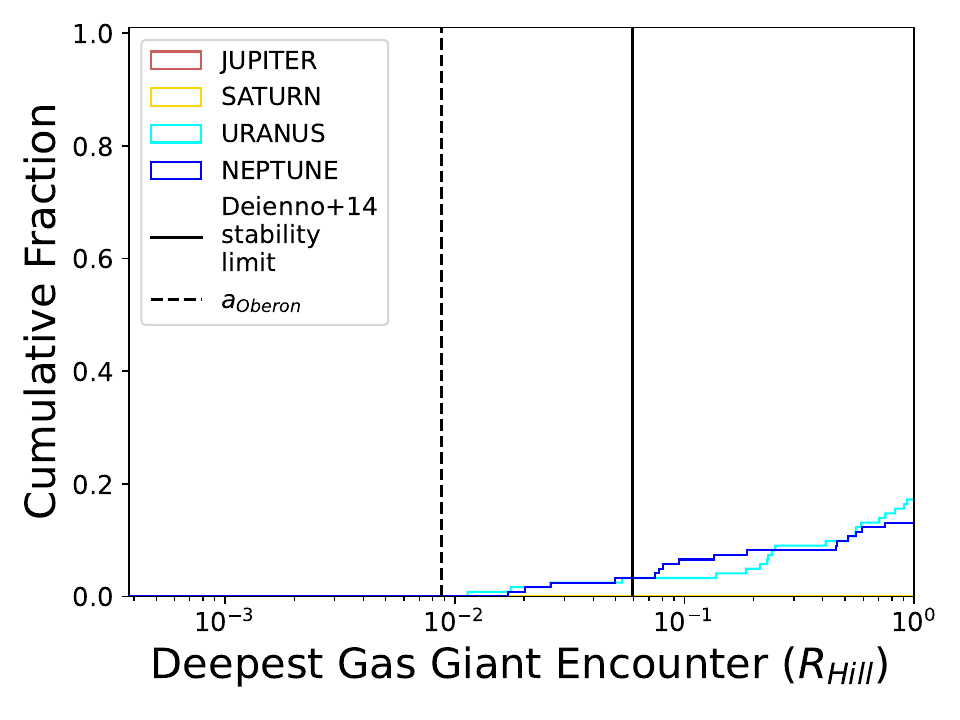}
    \caption{Cumulative distribution of the deepest encounter between one of the giant planets (red, gold, cyan and blue curves) and an Ice Giant (left panel) or Gas Giant (drawn from the combined distributions of encounters with Jupiter and Saturn, right panel) in our sample of 122 instability evolutionary tracks from \protect\citet{clement21_instb,clement21_instb2} that satisfied all important outer solar system dynamical constraints.  The dashed line plots the semi-major axis of Oberon, and the right vertical line plots the result of \protect\citet{deienno14} that encounters within 0.02 au regularly destroyed the Galilean moon system.}
    \label{fig:denc_stats}
\end{figure*}

As a starting point for our current study, we selected 122 plausible close encounter histories for the giant planets out of a sample of nearly 10,000 instability simulations originally reported in \citet{clement21_instb} and \citet{clement21_instb2}.  Each simulation used the \textit{Mercury6} Hybrid integrator to integrate the primordial giant planets along with 1,000 $\sim$Pluto-mass Kuiper Belt Objects (KBOs) through the instability for 20 Myr.  In order to fully probe the range of possible encounter histories that are capable of placing the giant planets on their modern orbits in as comprehensive of a manner as possible, we selected systems that simultaneously satisfied multiple broad dynamical constraints \citep[e.g.][]{nesvorny12}:

\begin{enumerate}
    \item Each system must finish with exactly 4 giant planets
    \item The final ratio of Saturn:Jupiter's orbital period, 2.2$<P_{S}$/$P_{J}<$2.8 (the current ratio is 2.49).
    \item Neptune's semi-major axis, $a_{N}<$ 40 au (30.1 au currently)
    \item The magnitude of Jupiter's fifth eccentric mode, $e_{55}>$ 0.022 (the modern value is 0.044)

\end{enumerate}

Of note, the excitation of $e_{55}$ is particularly important because satisfying it typically necessitates a number of reasonably strong encounters between Jupiter and other ice giants.  In many realizations, encounters at Jupiter are systematically weaker than at Saturn, and Jupiter's eccentricity is excited solely via eccentric forcing from an over-excited Saturn (thus Jupiter's sixth mode, $e_{56}$ is over-excited and $e_{55}$ is under-excited). 

\begin{figure*}
    \centering
    \includegraphics[width=0.49\linewidth]{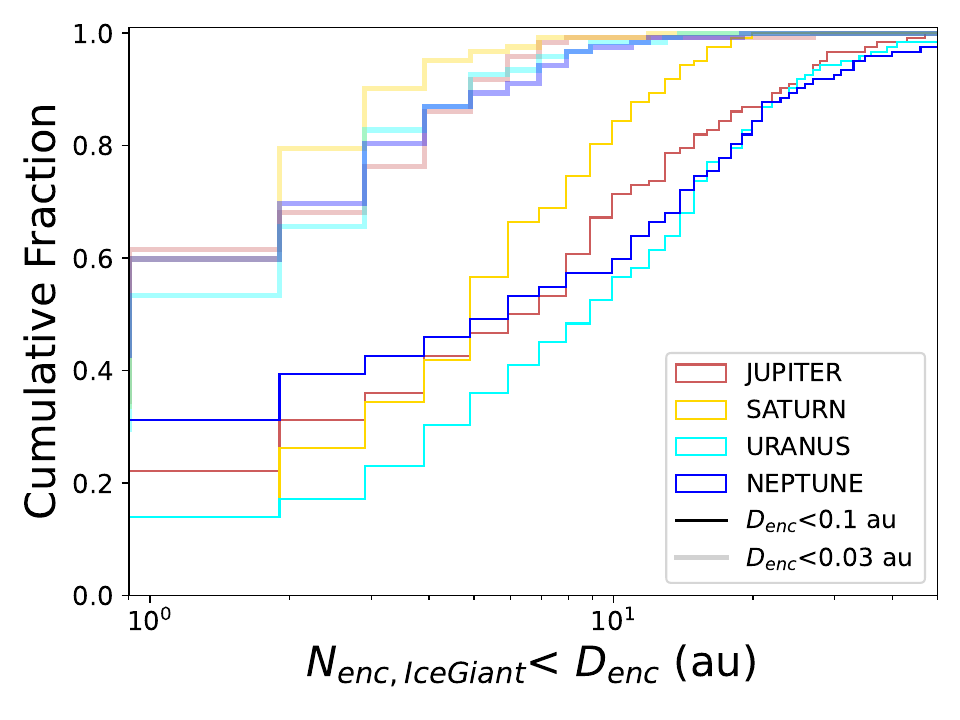}
    \includegraphics[width=0.49\linewidth]{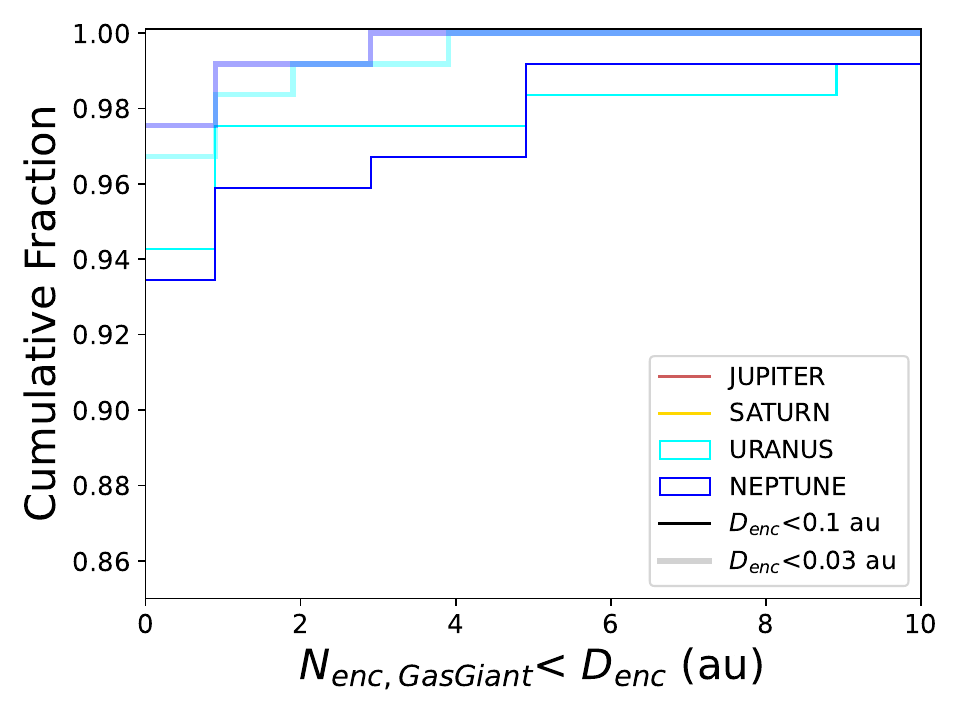}
    \caption{Cumulative distribution of the total number of successive encounters less than 0.1 au (solid lines) or 0.03 au (shaded lines) experienced by each giant planets (red, gold, cyan and blue curves) with an Ice Giant (left panel) or Gas Giant (right panel) in our sample of 122 instability evolutionary tracks from \protect\citet{clement21_instb,clement21_instb2} that satisfied all important outer solar system dynamical constraints.  Note that the x-axis in the left panel is log-scaled, while the right panel is not.}
    \label{fig:nenc_stats}
\end{figure*}

To boost the likelihood that the system finishes with four total giants and also minimize the time that resonances capable of disrupting Mars' orbit or over-exciting those of Earth and Venus linger in the vicinity of the terrestrial planets \citep{brasser09}, the current version of the Nice Model invokes the formation of an additional one or two primordial ice giants that are subsequently lost via dynamical scattering events during the instability \citep{nesvorny11}.  The original simulations from \citet{clement21_instb,clement21_instb2} varied a number of parameters including the total number of initial ice giants, the primordial resonant chain that the planets inhabit (in particular 35 out of our 122 cases begin with Jupiter and Saturn in the 3:2 resonance, and the rest begin with them in 2:1), the initial eccentricities of the planets, the spacing between Neptune and the primordial Kuiper Belt, and the mass of the primordial belt.  For the purposes of our current study, the most relevant varied parameter is the initial ice giant number.  45 of our selected systems contain one extra ice giant at time zero (henceforward referred to as ``5GP'', or five giant planet systems), and the other 77 possess two (henceforward ``6GP'').  This is important because the masses of the additional ice giants in our 6GP systems (6-8 $M_{\oplus}$) are lower than those in our 5GP system (12-16 $M_{\oplus}$).  Thus, the encounters between Uranus and these additional ice giants are gravitationally weaker at a given $D_{enc}$, but also more numerous in the 6GP portion of our sample than in the 5GP set.  We expound upon the implications of this throughout this manuscript, and we also briefly discuss the implications of the initial compactness of the resonance chain in section \ref{sect:ur_results}.

Figure \ref{fig:denc_stats} plots the distribution of the deepest encounters in each simulation between each planet and an ice giant (left panel) or a gas giant (right panel).  It is not surprising that Uranus and Saturn experience systematically deeper encounters ($\sim$5-15$\%$) than Jupiter and Neptune given their position in the system (i.e. they can experience close approaches with both exterior and interior planets.  It is also not surprising that deep gas giant encounters are fairly rare for Uranus and Neptune, and non-existent for Jupiter and Saturn.  In the latter case, if Jupiter and Saturn experience a close approach, the solar system is extremely likely to enter the scattering regime \citep{raymond09_scat,raymond10}.  In such evolutionary tracks, all ice giants tend to be lost in rapid succession, and the instability ends with Jupiter and Saturn as the lone survivors on well spaced, highly eccentric orbits \citep{deienno17}.

If we take the results of \citet{deienno14} for Galilean satellite stability during the instability as a rough starting point (solid black line in figure \ref{fig:denc_stats}), we would conclude that there is about a 50-60\% chance that the primordial satellite system of any of the four giant planets might have been destabilized by the giant planet instability.  At first glance, this finding might not appear particularly problematic.  Indeed, as discussed in section \ref{sect:intro}, it has been proposed that the satellite systems of Saturn and Neptune have each experienced at least some reorganization since the dispersal of the Sun's gas disk.  However, \citet{deienno14} also found that repeated chains of encounters at $D<$ 0.03 au are also capable of readily destabilizing the moons.  Figure \ref{fig:nenc_stats} depicts the total number of encounters at $D_{enc}<$ 0.03 au and 0.1 au in our sample of evolutionary tracks.  As repeated deep encounters are relatively common (for instance, $\sim$20\% of our cases have 5 or more such encounters at Uranus), we can reasonably expect the fraction of cases that destabilize the moons will be somewhat more than the $\sim$50\% estimated above based on the deepest encounter distance in each run.

We also note that, on the whole, our sample is largely consistent with the smaller sample of 3 instability evolutionary tracks tested by \citet{deienno14}.  The deepest encounters at Jupiter in their three cases were 0.01, 0.02 and 0.06 au, which fall at around the 30th, 50th and 70th percentile of the distribution for Jupiter in figure \ref{fig:denc_stats}.  Similarly, two of the three cases exhibited two total encounters with $D<$ 0.03 au, and the other case experienced none.  In figure \ref{fig:nenc_stats} we see that 35\% of our evolutionary tracks contain no such encounters at Jupiter, and another $\sim$35\% have either 1 or 2. 

\subsection{Satellite stability simulations}

We used a modified version of the $Mercury6$ Hybrid integrator originally described in \citet{kaib18} to model the stability and evolution of the Uranian moons during our selected epochs of planetary encounters.  The code has been extensively tested and employed in a number of prior works studying a variety of dynamical problems \citep{kaib18,ellithorpe22,kaibandraymond24,izidoro25_capture}.  We employ it here because it can accurately integrate systems with an arbitrary number of perturbing, passing bodies that have masses comparable to, or in excess of that of the central body.  The largest masses (for us the encountering objects) are integrated in inertial center-of-mass coordinates with a T$+$V leapfrog scheme, while lower mass bodies (moons) in orbit about the central body (Uranus and Jupiter in our models) are still integrated with a democratic heliocentric mixed variable symplectic algorithm \citep{wisdomholman,duncan98}.  While the code was originally developed to study the effects of passing stars on objects that orbit stars in binary systems, we were able to easily re-purpose it for our study with only a few changes.

All of our simulations include the central planet, its regular moons, and the Sun on their modern orbits (we assume that Uranus already attained its modern obliquity at the time of the instability).  We use a 0.07 day timestep, include the effects of Uranus' oblateness via its $J_{2}$ moment \citep{duncanandlissauer97,deienno11}, treat the Sun and all encountering objects (planets or KBOs) as T$+$V class particles, and integrate each system for 100 Myr.  

Our code accepts a list of encounter times, positions and velocities as an input. Each encountering object is then integrated until it exceeds its initial distance from the central body, at which time it is removed from the simulation.  Thus, we must first integrate the orbits of the encountering objects that were output to log files in the original simulations of \citet{clement21_instb,clement21_instb2} backwards in time to the point that we wish to inject the object into our full stability simulation.  Replicating the exact encounter geometry outside of Uranus' Hill sphere with this methodology is not possible without repeatedly stopping the simulation to change the position of the Sun, as the Uranus-Sun distance can change substantially throughout the initial instability simulation.  In the most extreme example, Uranus could have an encounter near perihelion with Jupiter around $\sim$5 au early on in the evolution that raises its semi-major axis.  Then, later in that same simulation Uranus might have an encounter with Neptune or the ejected ice giant near aphelion at $\sim$25 au.  On the other hand, if were to only model the encountering planets within Uranus' Hill Sphere, these additional complexities can be neglected.  To verify that this simplification would not affect our results, we performed a series of preliminary simulations varying the size of the sphere around Uranus in which the encountering planets were modeled (1, 2, 3, and 5 $R_{Hill}$) and determined that they all yielded similar results in terms of the evolution of the regular satellites.  In light of these results, we elected to generate our encounter initial conditions for our production runs by using an integration of the Sun, Uranus and the encountering object backwards in time from the point of closest approach as output in the original instability simulation, to separation of $D=R_{Hill}$.  We extensively verified that, using this approach, the \citet{kaib18} integrator reproduced the encounter distance and the angle between the encountering objects' velocity vectors within 0.1\% of the values reported in the original instability simulation.  

\subsubsection{KBO encounters}

\citet{deienno11} found that KBOs with masses $\gtrsim$ 0.15 times the mass of Pluto encountering within the orbit of Oberon can also excite the orbits of the regular moons well in excess of their modern values.  As including additional encountering objects is not particularly limiting computationally speaking (the most expensive part of our simulation is integrating the moons' orbits for \num{5e11} timesteps), we also performed simulations including the KBO encounters reported in the original instability simulations. In general, the fraction of destabilized satellite systems resulting from these simulations was nearly identical to the cases that only included planetary encounters.  This does not mean that KBO encounters are unimportant, or that they do not affect the evolution of the regular satellites at all.  Rather, it is simply a reflection of the fact that the vast majority of our simulations contained an overwhelming number of strong planetary encounters (figure \ref{fig:nenc_stats}). In this paper, we will focus on the effects of planetary encounters in isolation (i.e., our simulations that do not include KBO encounters).  In a follow-on manuscript, we will elaborate on the ways KBO encounters affect the evolving satellite systems.

\subsubsection{Effects of Tides}

\begin{figure}
    \centering
    \includegraphics[width=.99\linewidth]{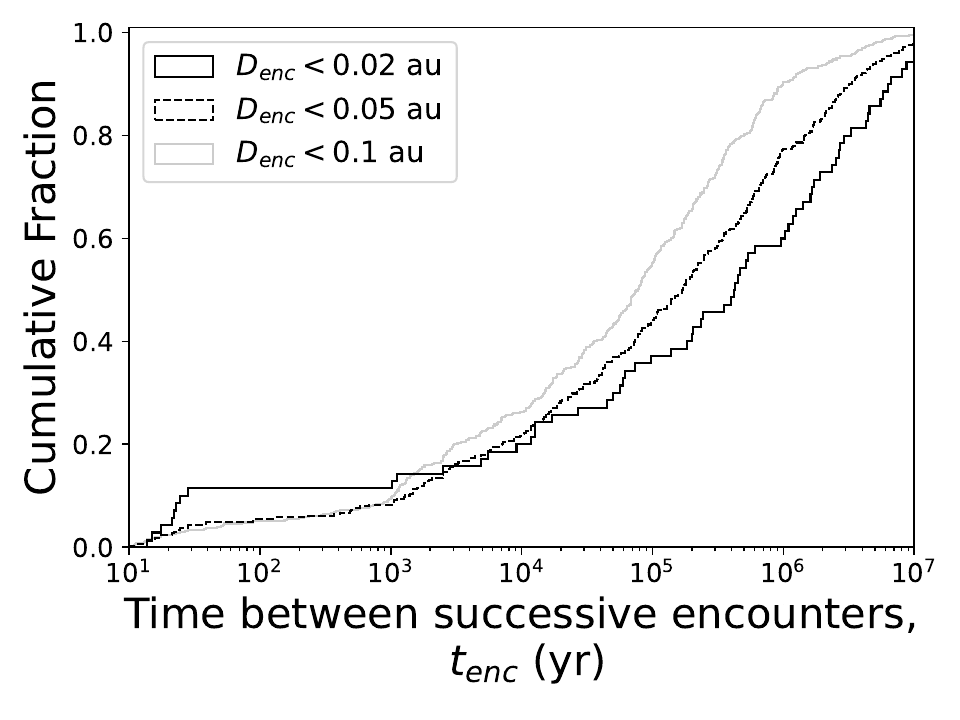}
    \caption{Cumulative distribution of times between successive encounters ($dt_{enc}$) less than various distance thresholds ($D_{enc}$) at Uranus in our 122 instability evolutionary tracks.}
    \label{fig:tenc_distrib}
\end{figure}

Tides are not expected to operate at a sufficiently rapid timescale (in terms of the moon's $\dot{a}$ and $\dot{e}$) to rapidly damp the orbits of a destabilized system and ``save'' it from experiencing collisions. Indeed, we observed many instances in our sample of encounter simulations where a planetary flyby with $D_{enc}\sim$ 0.02-0.05 excites satellite eccentricities to $\sim$0.05-0.10 without destabilizing them.  Then, a subsequent, moderately deep encounter further excites the moons to the point where they collide with one another.  While it is possible that, in some cases, neglecting tides might lead us to conclude that a system is unstable when, in actuality, the moons' orbits sufficiently damp between successive encounters such that they survive the sequence, these cases are likely quite rare.  Figure \ref{fig:tenc_distrib} plots the distribution of times between successive planetary encounters of different depths at Uranus in our sample of instability evolutionary tracks.  In the majority of cases, the time between pairs of encounters is less than about 1 Myr \citep[consistent with the results of statistical studies of the instability:][]{nesvorny12,deienno17,clement21_instb}, with instances of a pair of encounters book-ending a $\sim$10 Myr quiescent period being exceedingly rare.  

For an eccentric satellite in synchronous rotation with its host planet, the eccentric damping timescale is given by \citep[see section 4.10 of][]{dermott99}:
\begin{equation}
    \tau_{e} = -\frac{e}{\dot{e}} = \frac{2 m_{s}}{21 m_{p}} \bigg( \frac{a_{s}}{R_{s}} \bigg)^{5} \frac{Q_{s}}{k_{2,s}n}
    \label{eqn:edamp}
\end{equation}
 Here, the subscripts ``s'' and ``p'' denote values for the satellite and planet, respectively, and $n$ is the satellite's mean motion. All of the simulations in this paper follow the nominal model of \citet{cuk20} and assume values of 20,000 for Uranus' tidal $Q$ \citep[however, recently][found $Q=$678 for Uranus]{jacobson25} and 100 for each of the moons.  Similarly, we used tidal Love numbers of $k_{2}=$ 0.1 for Uranus, 0.001 for Miranda and 0.01 for the other four moons.  For the inner Uranian moons, assuming icy compositions, equation \ref{eqn:edamp} gives values of $\dot{e}$ of order \num{1e-4} units of eccentricity per Myr for moderate initial eccentricities of $\sim$0.05-0.10.  For the outer moons, values of $\dot{e}$ are closer to \num{1e-6}.  In our case, the slow damping of Oberon and Titania is most problematic as, once excited to eccentricities of order $\sim$0.10, their excitation continues to bleed throughout the system via strong secular forcing for 100s of Myr.  Thus, we do not expect tidal damping to be sufficient to prevent destabilization in the vast majority of our evolutionary tracks \citep[note, however, that $Q$ can change appreciably for high-eccentricity satellites as tidal heating melts the interior, and we do not include an eccentricity-dependent $Q$ in our models, e.g.][]{dermott99}.

We performed a small number of simulations with tidal forces included to more concretely justify our choice of neglecting the effect in our production simulations.  We accomplished this by following the classic, constant time-lag model of \citet{mignard79} and applying an artificial force on each moon's orbit \citep[see also:][]{bolmont15}.  We verified our implementation of tides in the \citet{kaib18} version of $Mercury$ through a series of 100 Myr integrations of the modern system.  We tested a variety of different sets of orbital configurations for the Uranian system (in terms of initial inclinations and eccentricities), and verified that the orbital migration and eccentricity damping rates ($\dot{a}$ and $\dot{e}$) were largely consistent with the analytically calculated values \citep{goldreich66,dermott99}, and those presented in previous numerical studies of system utilizing our same values of $Q$ and $k_{2}$ \citep[specifically,][]{cuk20}.  We did not identify any cases where tidal damping of eccentricities prevented an instability that transpired in one of our non-tidal cases from occurring.

\subsection{Full instability simulations}

The results presented in the subsequent sections comprise 1,342 total simulations of Uranus' moons: 10 sets of 122 distinct instabilities without tides, and one set with tides. To generate unique simulations, the satellites orbits are initialized in different orbital phases in each individual run.  We also ran one simulation for each encounter history focused on Jupiter's moons with an identical setup (122 total simulations) for the purposes of making a better comparison with \citet{deienno14}, and to quantify how much more sensitive the Uranian system is to encounters during the instability than the Jovian system.  The majority of our simulations were run on the Texas Advanced Computing Center’s (TACC) \textit{Frontera} cluster.  The median runtime for an individual simulation was 1292 hours, a little less than 2 months.  The longest run took 1739 hours, closer to 3 months to complete.  With the addition of 122 Jovian simulations, the total sample of complete Nice Model evolution simulations presented in this paper is 1,464.

We found that only 13\% of our 1,464 satellite systems remained stable for the duration of the simulation (i.e. did not lose a moon via collision or ejection).  The majority of these instabilities ensue quite rapidly after particularly deep encounter.  We discuss the dynamics of some of the instabilities themselves in sections \ref{sect:5gpvs6gp} and \ref{sect:consequences}.  Our fraction of stable systems is far less than the $\sim$40-50\% stability rate we naively expected when comparing the distribution of deepest encounter depths (figure \ref{fig:denc_stats}) with the results of \citet{deienno14}.  A major reason for the discrepancy turned out to be the effects of repeated encounters at moderate depths (see additional discussion in section \ref{sect:follow-on-results}). To verify this, we performed an additional number of encounter simulations, described in the subsequent section.

\subsubsection{Additional simulations investigating the effects of repeated encounters}
\label{sect:follow-on}

Our sample of encounter histories contain many more runs with large numbers of repeated encounters at moderate depths (figure \ref{fig:nenc_stats}) compared to the evolutionary tracks evaluated in \citet{deienno11} and \citet{deienno14}.  To quantify how this affects our results, we conducted a follow-on suite of many shorter simulations utilizing fictitious encounter histories and the same modeling framework described above.  We begin by compiling three libraries of all planetary encounters with each respective giant planet with 0.008 $<D<$ 0.015 au, 0.04 $<D<$ 0.06 au and 0.09 $<D<$ 0.11 au by drawing from our complete suite instability evolutionary tracks.  In the subsequent text, we refer to these encounters simply as having depths, $D_{enc}$ of 0.01, 0.05 and 0.1 au.  In each simulation we randomly select $N_{enc}$ encounters with $D_{enc}$, space them in time by $dt_{enc}$, and integrate the entire system for 10 Myr after the final encounter.  We tested values of  $N_{enc}=$ 5, 10 and 20; and $dt_{enc}=$ 100, 1,000 and 10,000 years.  For each permutation of $D_{enc}$, $N_{enc}$ and $dt_{enc}$ we performed 100 integrations using different initial phasings of the satellite system.  We did not consider any gas giant encounters with Uranus at 0.01 au, or Jupiter-Saturn encounters as we know they will readily destabilize the system.  Thus, we performed 4,500 simulations focused on the Uranian system ($D_{enc}=$ 0.01, 0.05 and 0.1 au for ice giant encounters,  $D_{enc}=$ 0.05 and 0.1 au for gas giant encounters\footnote{Recall that Uranus only experiences an encounter at $D_{enc}<R_{Hill}$ with one of the gas giants in about 20\% of our sample of instability simulations, see figure \ref{fig:denc_stats}.}, and 9 permutations of $N_{enc}$/$dt_{enc}$) and 2,700 simulations of the Jovian system (the same parameter space without gas giant encounters).  We discuss the results of these additional simulations in section \ref{sect:follow-on-results}.

\section{Results and Discussion}

\subsection{Satellite system survival}
\label{sect:ur_results}

\begin{figure}
    \centering
    \includegraphics[width=.99\linewidth]{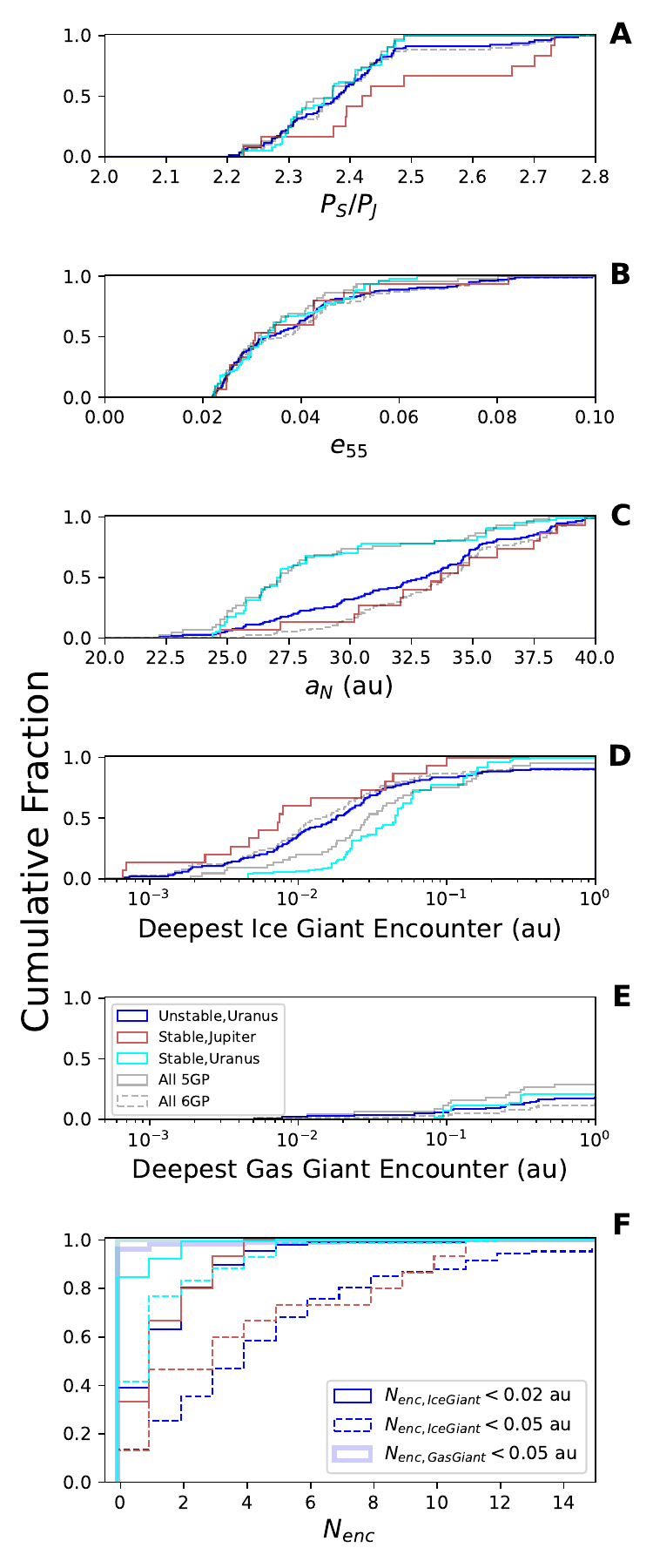}
    \caption{Cumulative distributions of various properties of instability evolutionary tracks where the regular satellites remained stable (light blue for Uranus and red for Jupiter) or were destabilized (dark blue for Uranus).  For comparison, we also plot the distributions for all 5GP (solid grey line) and 6GP (dashed grey line).  Panel F shows the total number of Ice Giant encounters deeper than 0.02 and 0.05 au, as well as Gas Giant encounters deeper than 0.05 au for these same collections of simulations.}
    \label{fig:all_stats}
\end{figure}

We consider a system to be destabilized when a pair of satellites collides, or if a satellite is removed from the system via ejection from its planetocentric orbit, scattered onto an orbit with a semi-major axis in excess of 2$a_{Oberon}$ (Oberon's semi-major axis), or merger with the central planet.  In reality the overwhelming majority of cases of instability result in collisions between satellites.  Overall, 87.3\% of our simulations of the Uranian moons destabilized.  Similarly, 87.7\% of our simulations of the Galilean satellites yielded instabilities, and only 3.3\% (4 simulations) survived with the Laplace resonance intact \citep[we still view those that perturbed the resonance as viable sequences for the solar system since subsequent tidal migration can reconstitute the resonant chain:][]{greenberg87,deienno14,fuller16}.  If we instead tabulate how many realizations of each distinct instability evolution destabilize Uranus' moons\footnote{Recall that we run each evolution 11 times at Uranus using different initial orbital phasings for the moons}, we find that 98.4\% (all but 2 instabilities) did so least once, 91.0\% did so at least 50\% of the time, and 58.2\% did so every single time.  

The probability that both Uranus and Jupiter's moons survive the same instability is only about 1\%.  Specifically, one of the stable Jupiter cases also yielded 6/11 stable Uranian simulations, and a second instability produced 1/11 stable Uranian simulations.  In the former (most successful) case, Jupiter experiences 22 encounters at $D_{enc}<$ 0.1 au; the deepest 4 of which have $D_{enc}\simeq$ 0.03 au.  As a result, the Laplace resonance is totally broken and Europa's eccentricity is excited to $\sim$0.08.  Uranus, on the other hand, only experiences 6 such encounters, the deepest being one with $D_{enc}=$ 0.057 au.  One obvious reason this sample of instabilities that allow for the simultaneous survival of both satellite systems is so small is that it restricts the Jovian encounter histories to not just those that avoid deep encounters with the ejected ice giant, but also those that avoid wider ($D\sim$ 0.1 au) encounters with Uranus.  When we couple these results with the encounter statistics of figure \ref{fig:nenc_stats}, we conclude that it is highly unlikely that the primordial satellites of at least one or more of the four giant planets were not strongly perturbed to the point of destabilization by the giant planet instability.  While the presence of the Laplace resonance likely places a constraint on the strength of encounters at Jupiter during the real instability, it is not implausible that the Uranian moons experienced some late reorganization after their formation.  In section \ref{sect:consequences} we will speculate that the most likely scenario for the solar system is one where the instability moderately perturbs the resonant chain around Jupiter\footnote{Note that, unlike \citet{deienno14}, we do not test cases where Callisto is initialized in resonance with the other 3 moons}, and causes some degree of consolidation or mass transfer between Uranus' primordial moons.

In general, stable realizations at Uranus possess no ice giant encounters $\lesssim$ 0.02 au, and do not include encounters with either Jupiter and Saturn within about 0.1 au.  Figure \ref{fig:all_stats} plots several properties of the instabilities where the Uranian moons remained stable, compared to those where the system was destabilized.  We also compare these results to the outcomes of our smaller sample of simulations focused on the Gallilean moons (red lines).

Unsurprisingly, the strongest predictor of stability is the number and depth of the encounters in the instability evolution (panels D and F in figure \ref{fig:all_stats}).  The initial number of ice giants also correlates with stable evolutions at Uranus (5GP) and Jupiter (6GP) in many of the panels.  We discuss these trend in detail in section \ref{sect:5gpvs6gp}.  The final configuration of Jupiter and Saturn -- in terms of their separation, $P_{S}/P_{J}$ and excitation (e.g. $e_{55}$) -- is not a strong determiner of whether either satellite system will remain stable.  Uranus' moons are slightly less stable at higher values of $e_{55}$ (particularly those approaching and in excess of the real value of 0.044), however this affect is minor.  We confirmed this by performing a Kolmogorov-Smirnov test on the two distributions that yielded a p-value of 0.048.  Moreover, our sample contains multiple examples of both stability, and destabilization of either satellite system for instabilities that almost perfectly replicate the solar system values of $P_{S}/P_{J}$ and $e_{55}$.

The final location of Neptune (panel C in figure \ref{fig:all_stats}) also appears to be somewhat predictive of stability for Uranus.  Here, we are essentially using Neptune's position as a proxy for the level of ``compactness'' of the giant planet system following the instability, as the final location of Jupiter in our entire sample only varies between 5.13-5.55 au.  Stable sequences almost exclusively occur when the giant planet's orbits are significantly more compact than they are today.  These instabilities are typically weaker and shorter lived than those where Neptune and Uranus' final semi-major axes are closer to the real values.  They are characterized by strong, decisive encounters at Jupiter and Saturn, and a very brief episode of encounters among the ice giants (contrast, for instance, the encounter tracks for Uranus depicted in figure \ref{fig:5gpvs6gp}; discussed further in section \ref{sect:5gpvs6gp}).  The longer the epoch of intra-ice giant encounters lasts, the more likely they are to experience approaches close enough for satellite destabilization and large changes in planet semi-major axes that tend to push Neptune beyond its current orbit.  

\begin{figure}
    \centering
    \includegraphics[width=.99\linewidth]{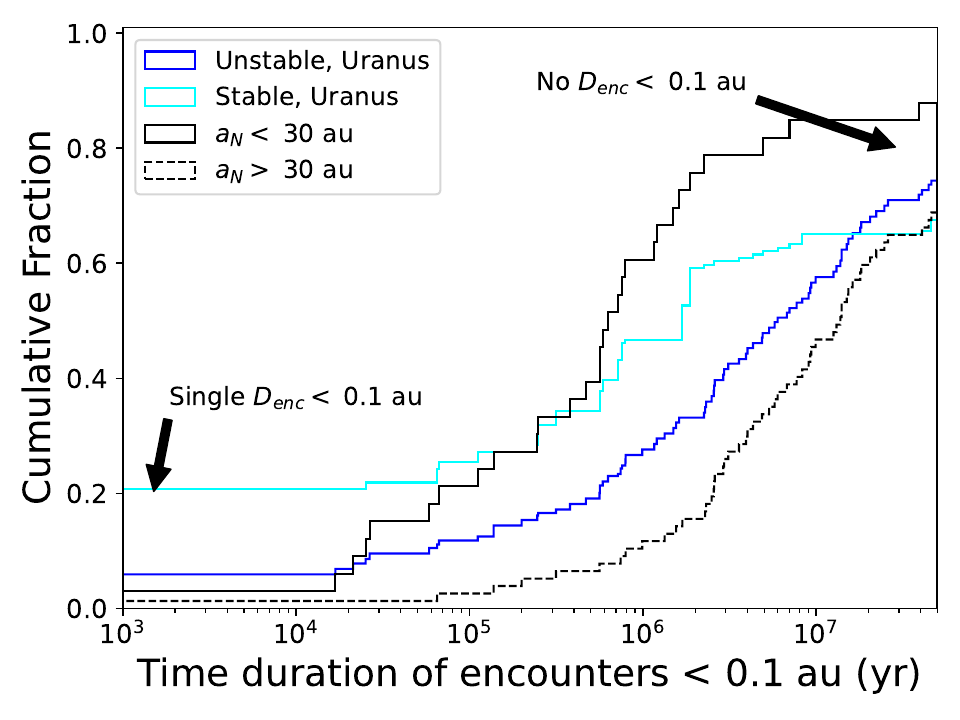}
    \caption{Cumulative distribution of duration of the times between the first ice giant encounter at Uranus with $D_{enc}<$ 0.1 au, and the last such encounter.  The different shades of blue lines show instabilities where the Uranian moons survived, and those where they did not.  The different styles of black lines show instabilities where Neptune finished inside of, and outside of its modern semi-major axis.  Here, we plot systems with only a single encounter with $D_{enc}<$ 0.1 au at 0, and those with no such events at infinity.}
    \label{fig:delta_tenc}
\end{figure}

Figure \ref{fig:delta_tenc} demonstrates how the final degree of compactness of the giant planet system (again we are using Neptune's final semi-major axis, $a_{N}$, as a proxy) correlates with the length of the epoch of deep encounters during the instability.  While each instability is necessarily unique and complex (e.g. different numbers of initial ice giants, different starting locations of Neptune, etc.), the overall relationship between the duration of the encounter epoch of the instability and the fate of the ice giants is quite clear.  Instabilities that tend to ``drag out'' -- with the extra ice giants spending more than a few Myr on planet-crossing orbits before ultimately being removed via ejection -- are more likely to destabilize satellites and produce a final system of ice giants with orbits that are too dispersed.  In contrast, when the instabilities are brief, in many of the cases Uranus simply never gets a chance to end up at the ``wrong place at the wrong time'' such that it undergoes a particularly close encounter before all of the action is over.

Given the small number of surviving satellite systems in our full simulation suite, we don't view this result as evidence that Uranus' moons survived the solar system's real instability simply because Neptune actually ended up at 30 au.  Indeed, the majority of our realizations where the moons survive come in cases where the giant planets finish on \textit{more compact} orbits than their modern ones ($a_{N}\sim$25-27 au).  If this were the case, a significant planetesimal population of bodies in the Kuiper Belt would have to survive the instability in order to pull the ice giants an additional 3-5 au through planetesimal-driven migration \citep{fernandez84}.  This kind of excessive post-instability migration is at odds with Kuiper Belt formation models \citep[e.g.][]{nesvorny15a,nesvorny16,kaib24} that tend to favor a more prolonged epoch of pre-instability migration, a short and abrupt change in semi-major axis during the instability, and a post-instability migration phase of less than a few au\footnote{Note that the value we report as Neptune's semi-major axis includes several 10s of Myr of residual migration, depending on when the instability actually occurred in the simulation.}.  Thus, the majority of our simulations that yield stable satellite sequences at Uranus are not the best solar system analogs because Neptune does not undergo significant pre-instability migration, and the giant planets' final orbits are too close to one another.  Moreover, when we down-select to the 16 total instabilities that finished with 29.0 $<a_{N}<$ 31.0 au, 13 destabilized the Uranian moons in every single case, 2 did so in the majority of our realizations, and only one consistently allowed for satellite survival (only one of our 11 simulations with that case yielded a satellite instability).  In the most successful case, however, Neptune migrated nearly 5 au via residual migration after the instability, which is in excess of the 2-3 au that is preferred by Kuiper Belt formation models \citep[e.g.][]{nesvorny15a}.  The most successful evolution in terms of simultaneously producing the preferred migration sequence for Neptune and allowing for Uranian satellite survival was one where Neptune migrated to 26.3 au before the instability, finished at 28.3 au, and Uranus' moons survived in all 11 of our realizations.  However, the issue remains that when we down-select to the systems with the best Neptunian migration histories, the survival probability for Uranus' moons remains close to 10\%.

In summary, we did not find any particular combination of qualities of the post-instability outer solar system that drastically improved the chances of satellite survival for Uranus.  The fundamental problem remains that planetary encounters at Uranus are too numerous and too deep in the vast majority of our simulations (figure \ref{fig:nenc_stats}).  Even in more rapid, well-behaved instabilities that don't overshoot Neptune's semi-major axis, both Uranus and Neptune almost always experience encounters with other ice giants within 0.05 au.  What is more predictive of encounters is the manner in which Uranus' orbit evolves within the instability, rather than the orbit it finishes the event on.  If Uranus' eccentricity is higher, or moderately excited for a longer period of time during the instability, it is in turn more likely to experience excessive deep encounters capable of destabilizing its satellites.  It is worth pointing out that, in some realizations of the Nice Model, Uranus and Neptune actually exchange positions.  This was first noticed to occur in $\sim$50\% of the initial simulations of \citet{tsiganis05}.  This behavior is present in 42 of the 122 instabilities that we study here.  As a result of the systematically larger eccentricities attained by the ice giants in these cases, along with the increased time Uranus and Neptune spend on crossing orbits, the ``swap'' style instabilities are modestly more likely to destabilize Uranus’ moons.  The average number of Uranus-Neptune encounters at $D_{enc}<$0.05 au is 3.83 in these cases, as opposed to 1.85 in the 82 models where the ice giants don't change position.  Similarly, 60\% of the non-swapping sample possess no Uranus-Neptune encounters within 0.05 au, as compared to only 24\% of the swapping models.  As a result, 71\% of the swapping cases destabilized Uranus' moons in all 11 simulations (as compared to 58\% of all simulations), and 95\% did so in at least 6 simulations (91\% for all simulations).


\subsubsection{Jovian satellite survival}

The story is the same at Jupiter: encounter depth and number governs stability.  However, the system is clearly more resilient to the perturbations of the encountering objects than the Uranian system.  Interestingly, the ``typical'' chains of encounters that are able to destabilize the moons of Uranus are quite similar to those that allow for Jovian system survival (dark blue and red lines in panels D and F in figure \ref{fig:all_stats}).  Indeed, Uranus' moons tend to readily destabilize after 1-4 encounters with $D_{enc}<$ 0.02 au or more prolonged chains of encounters closer to 0.05 au, while similar sequences at Jupiter can occasionally result in stability -- particularly if the encountering ice giants are less massive (discussed further in section \ref{sect:5gpvs6gp}).  Thus, though the most common outcome of any given instability in our sample is the destabilization of both systems of moons, survival at Jupiter and destruction at Uranus also occurs at a reasonable frequency.  We will elaborate more on this result in section \ref{sect:consequences}.

\subsubsection{Encounters between Uranus and the gas giants}

As discussed in section \ref{sect:methods1}, our sample of instabilities do not include any cases where Jupiter and Saturn encounter one another because those tend to result in scattering and poor solar system analogs.  About one quarter of our cases do include close encounters between Uranus and one of the gas giants, and those exchanges proved to be particularly problematic for Uranus' satellites.  The deepest encounter with a gas giant that allowed for Uranian satellite survival in any of our simulations was a 0.095 au exchange with Saturn.  Overall, this particular instability was fairly mild in terms of the depth of ice giant encounters with Uranus.  The deepest such events comprised a series of 3 moderate encounters (0.047, 0.056 and 0.090 au) with the ejected ice giant between $t=$ 4.5 and 4.8 Myr.  As discussed above, we find these types of ice giant encounter sequences are typical within the sample of instabilities where Uranus' moons survive.  However, things play out quite differently at Jupiter during this same time period as it experiences encounters at 0.038 and 0.024 au with the soon-to-be ejected ice giant that destabilize its moons.  

Given the scarcity of deep encounters between surviving ice giants and gas giants in our sample of instabilities that produced reasonable analogs of the solar system (figure \ref{fig:denc_stats}) -- coupled with the low probability of \textit{both satellite systems} surviving such an exchange -- we conclude that it is an unlikely that such an event played a relevant role in the early evolution of the solar system.

\subsection{The case for two extra ice giants}
\label{sect:5gpvs6gp}

Another reasonably strong predictor of satellite survival at Uranus and Jupiter is the total number additional ice giants in the initial instability evolution.  Recall that the addition of a fifth (5GP), or even sixth (6GP) additional giant planet to the primordial solar system is required to boost the probabilities of (1) finishing with four total giant planets and (2) triggering an abrupt change in Jupiter and Saturn's position with respect to one another that minimizes the effects of secular resonances dragging \citep{brasser09,nesvorny11} through the asteroid belt and across the orbits of the terrestrial planets \citep[though this could also be avoided if the instability occurred early:][]{clement18}.  

While 5GP cases makeup 36.9\% of our sample of tested cases, these evolutionary tracks yielded 70.2\% of our stable Uranian systems.  In contrast, while 6GP instabilities are more numerous in our sample \citep[quite simply because they more regularly yield good solar system analogs, see additional discussion in][]{clement21_instb,clement21_instb2}, only three of these 77 instabilities consistently\footnote{We consider a given instability successful in this manner if the satellites remained stable in at least 6 of the 11 simulations.} produced stable evolutions at Uranus.  We attribute this result to the increased number of deep encounters between ice giants in the six planet cases.  Indeed, the average number of ice giant encounters at $D_{enc}<$ 0.05 au in our 5GP sample is 2.87.  Unsurprisingly, the average for 6GP cases is 5.38, about twice as many.  Moreover, 31.1\% of our 5GP instabilities yield zero such encounters with Uranus, as compared to only 9.1\% of 6GP set.  Thus, it is more likely for the quick, well-behaved ideal type of evolution where Uranus fortuitously avoids encounters to occur in the 5GP sample.  

While the 5GP model present the best case scenario for Uranus, we find it to be significantly more problematic for Jupiter's satellites.  12 of our 15 stable evolutionary tracks at Jupiter (80$\%$) occurred in 6GP simulations.  While the total number of encounters is still greater in the six planet case, the typical depth at which they tend to occur ($\sim$0.01-0.1 au) is more survivable for Jupiter's moons if the encountering object is only around 8 $M_{\oplus}$\footnote{The additional two ice giants in our 6GP instabilities have masses between 6-10 $M_{\oplus}$, while the additional planets in the 5GP cases have masses between 12-20 $M_{\oplus}$.  Smaller additional ice giants tend to boost the probability of the surviving planets attaining the correct final semi-major axes in the 6GP case.  For reference, the three cases considered in \citet{deienno14} use 16 $M_{\oplus}$ ice giants.  However, it is not the case the 6GP dynamical histories evaluated here simply contain twice as many encounters with planets that are half as massive because each set is drawn from a small sub-sample of cases that best match the solar system.  The types of 5GP and 6GP sequences that best reproduce the modern solar system are characteristically different.  Specifically, in the 6GP set, Jupiter and Saturn's orbits are excited over a much more prolonged sequence of moderate encounters with the additional planets.  Consequently, the extra ice giants spend lengthy times on crossing orbits with Uranus.}.  In contrast, encounters at this depth at Uranus are still destructive if the encountering ice giant is slightly smaller.

\begin{figure*}
    \centering
    \includegraphics[width=0.95\linewidth]{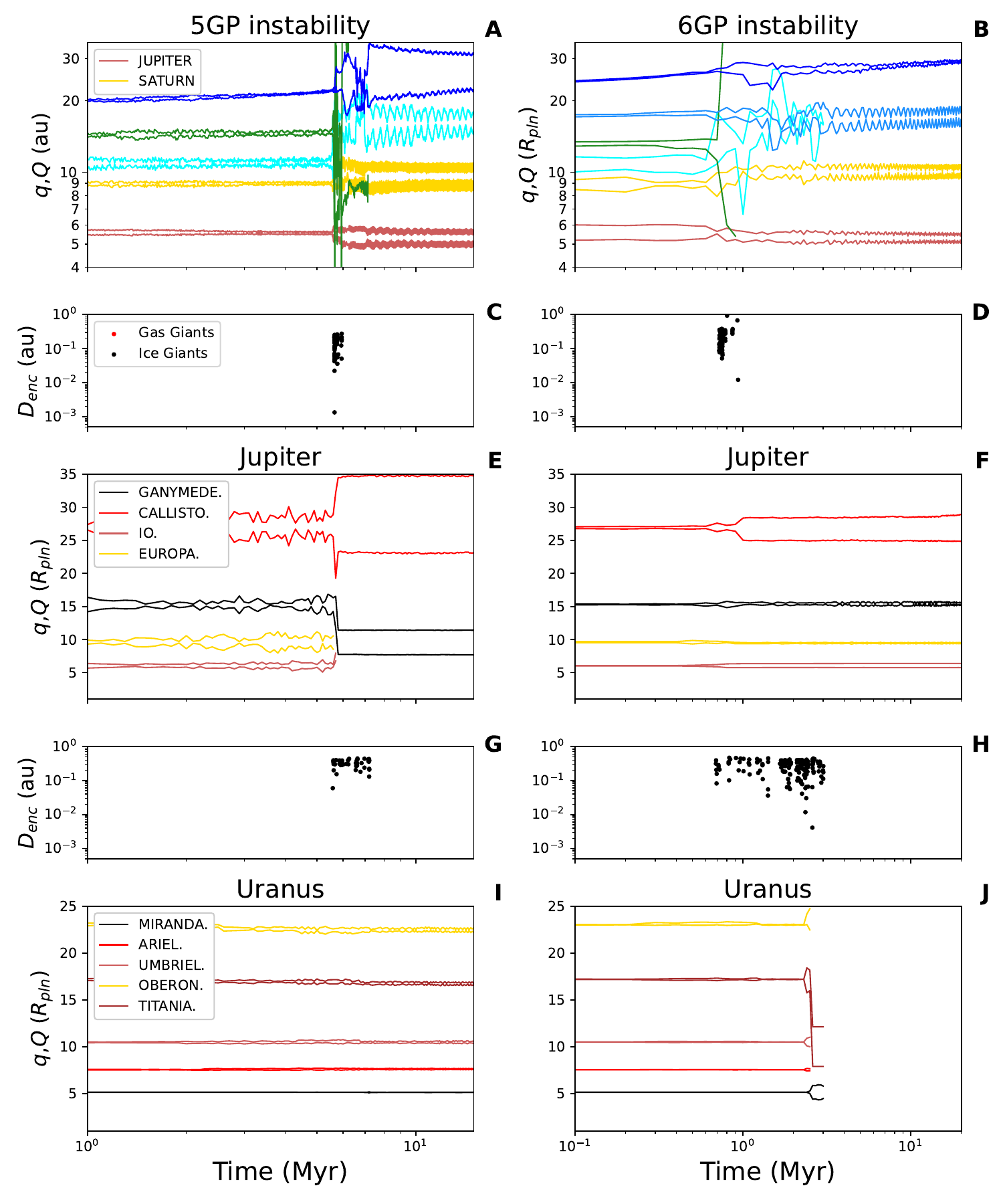}
    \caption{Comparison of a typical evolution beginning with five total giant planets (left panels) to one with six initial planets (right panels).  Panels A and B show the evolution of the perihelia and aphelia of each giant planet.  Panels C and D show the encounter histories at Jupiter, and panels E and F depict the orbital evolution of the Jovian moon's periapsis and apoapsis.  Panels G-J show the same data for the same evolutionary tracks for the Uranian system.}
    \label{fig:5gpvs6gp}
\end{figure*}

Figure \ref{fig:5gpvs6gp} compares a ``typical'' 5GP instability (left panels) where Jupiter's moons are destabilized and Uranus' survive, with a typical 6GP case that produces the opposite result.  In the 5GP case, Jupiter experiences 49 total encounters with the ejected ice giant within its Hill sphere (panels A and G) just before $t=$ 6 Myr.  In this case the ice giant that is lost is the one initialized with an orbit in between Uranus and Neptune.  The deepest exchange is well within the orbits of the Galilean satellites: 0.0013 au at 5.65 Myr.  This exchange kicks Jupiter, and the satellite system is destroyed in short order.  Io collides with Europa, and our code merges the two moons into a single body that collides with Ganymede 100 kyr later (panel E of figure \ref{fig:5gpvs6gp}). Panels G and I on the left side of figure \ref{fig:5gpvs6gp} show the consequences of the same instability evolution at Uranus.  During the $t=$ 5.6 Myr series of deep encounters with Jupiter, the ejected ice giant is on a crossing orbit with Uranus with a perihelion near Jupiter. This configuration produces a series of 3 close approaches with Uranus.  The deepest of these encounters is 0.059 au -- comfortably outside of the depth that we find regularly destabilizes Uranus' moons (figure \ref{fig:all_stats}).  After the 0.0013 au approach at Jupiter raises the extra ice giant's semi-major axis and perihelion, it experiences a chain of encounters with Saturn prior to its ultimate ejection between $t=$ 6 and 7 Myr.  While the ice giant is interacting with Saturn at perihelion, it remains on a crossing orbit with both Uranus and Neptune that leads to a series of 28 additional distant encounters with Uranus.  The deepest of these events is an encounter at 0.13 au at $t=$7.2 Myr. Thus, as a result of their fortuitous avoidance of close encounters, we find that Uranus' moons survive in all but one of our simulations.  In the single unstable phasing, the 0.059 au encounter at $t=$5.68 Myr moderately excites the eccentricities of all the moons to between 0.01-0.05.  In the aftermath of this encounter, Miranda's eccentricity continues to gradually excite over the subsequent $\sim$10 Myr until it collides with Ariel.

In the 6GP case depicted in the right panel of figure \ref{fig:5gpvs6gp}, Jupiter begins to interact strongly with the second of two additional, lower mass ice giants initialized between Saturn and Uranus around $t=$ 720 kyr.  The deepest of these 67 exchanges at 934 kyr with $D_{enc}=$ 0.012 au is responsible for the additional planet's ejection.  However, because the ejected planet only has a mass of 6.0 $M_{\oplus}$, its influence is strong enough to break the Laplace resonance and modestly excite the orbit of Callisto to $e=$ 0.2, but not strong enough to fully destabilize the system.  In contrast to the other three encounter evolutionary tracks depicted in figure \ref{fig:5gpvs6gp}, the six planet case at Uranus (panel H) is by far the most prolonged.  In total, Uranus experiences 160 total encounters with both ejected ice giants inside its Hill sphere (one of which is ejected by Jupiter and the other being ejected by Saturn).  20 of these encounters are within $D_{enc}=$ 0.1 au, 6 are inside of $D_{enc}=$ 0.05 au, and one occurs interior to $D_{enc}=$ 0.01 au.  The total number of encounters is hardly relevant though.  The first deep encounter at 0.01 au around $t=$ 1.3 Myr excites the outer moons and leads to collisions between Titania, Umbriel and Ariel.  An additional encounter with $D_{enc}=$ 0.003 au that kicks Uranus and its satellites at $t=$ 1.7 Myr is then responsible for the ejection of the remaining moons.

Because of the Laplace resonance, we view the survival of Jupiter's moons as a stronger constraint on the instability evolution than stability at Uranus.  While it is plausible that the resonance was briefly broken and reconstituted via tidal migration \citep[e.g.][]{greenberg87,deienno14,fuller16}, it is less clear whether a total break-up and re-accretion of the satellite system is compatible with constraints from the satellite's cratering records \citep{wong23,bottke24} and Io's lengthy history of intense volcanism \citep{dekleer24}.  As fewer constraints on the Uranian satellite's long-term dynamical history exist, when taken at face value our results can be interpreted as favoring a scenario where the primordial outer solar system possessed six total planets, rather than five.  While the trends observed in our simulations are compelling, they are by no means strong enough to totally rule out the 5GP version of the Nice Model.  Indeed, we observe multiple cases of Jovian survival with five planets, Uranian system survival with six planets, as well as six planet sequences with very brief epochs of deep encounters.  

Nevertheless, it is interesting and certainly worth pointing out that each of the two instabilities simultaneously survived by \textit{both} Jupiter and Uranus' satellites contained five initial planets.  Thus, the most likely pathway to satellite survival at both Jupiter and Uranus is a 5GP instability where Jupiter ejects the extra ice giant through an exchange around $D_{enc}\simeq$ 0.05 au after a relatively brief epoch of encounters, thus allowing Uranus to fortuitously avoid deep encounters entirely.  However, even our lone case (from a sample of 122 evolutionary tracks handpicked from a larger sample of almost 10,000 simulations) that proved somewhat viable on these grounds is far from perfect since it still breaks the full Laplace chain, over-excites the eccentricities of the Galilean moons and only resulted in survival of Uranus' moons in half of our trials.

\subsubsection{Effect of other initial conditions}

Our sample of instabilities can also be broken down into four groups of different resonance chains for Saturn and the first three ice giants; thus combining the sets of simulations initialized with different Jupiter-Saturn resonances, along with those in 5GP and 6GP sets by excluding the last resonance with Neptune.  The resulting partial chains -- starting with the resonance between Saturn and the first ice giant -- are: (1) 3:2,3:2,3:2, (2) 4:3,3:2,3:2, (3) 4:3,4:3,3:2 and (4) 4:3,4:3,4:3.  This breakdown groups our sample into four different levels of initial planetary compactness around Uranus.  We observe a clear trend of increasing rates of Uranian satellite instabilities with increasing degrees of compactness of the initial resonances around Uranus: (1) 78\%, (2) 83\%, (3) 91\% and (4) 97\%\footnote{Only 7 of our 122 instabilities are from initial chains containing three 4:3 resonances because they are less likely to match constraints for the outer solar system \citep[see section \ref{sect:methods1}, and additional discussion in][]{clement21_instb}, and therefore these results are more difficult to interpret in light of the small number statistics.}.  This is a result of the fact that deep planetary encounters are more likely to occur when the giant planet configuration is initially more compact.  As the probability of satellite survival is low for all configurations, this result does not particularly change the overall conclusions of our analysis.  As of the writing of this paper, strong constraints on the actual initial resonant chain in the outer solar system do not exist.  If such constraints became available in the future (for instance from giant planet formation of disk evolution models), they could be used in tandem with our results to place even stronger constraints on the encounter histories of each planet and their effects.

We were unable to identify any correlations between satellite survival at each planet and other qualities of the initial giant planet configuration such as the primordial resonance inhabited by Jupiter and Saturn \citep[3:2 or 2:1,][]{clement21_instb}, the initial eccentricity of the giant planets, or the mass of the primordial Kuiper Belt.  While many of these parameters can boost the probability of an individual initial configuration producing a good solar system analog, this is not particularly relevant to our study as we have already down-selected to the ``best'' analogs.   In conclusion our results lead us to favor the six planet instability over the five planet version because it produced lowest fraction of unstable systems (84\% at Jupiter) of any instability/planet combination in our study.



\subsection{How many encounters are too many?}
\label{sect:follow-on-results}

While destabilization via a single, deep encounter that kicks the planet is the most straight-forward sequence of events responsible for satellite system destruction, our sample of simulations also contains many evolutionary tracks where multiple encounters at moderate depths perturb one of the outer satellites and are thus responsible for producing an instability within a moon system.  The six planet example at Uranus in figure \ref{fig:5gpvs6gp} (panels H and J) is one such example of this behavior.  In that case, the instability ensues precipitously after a series of moderately deep encounters. In order to quantify the degree to which the number, depth and temporal spacing of consecutive encounters affects satellite system stability, we performed an additional set of simulations using contrived sequences of $N_{enc}=$ 5, 10 and 20 encounters at depths of $\sim D_{enc}=$ 0.01, 0.05 and 0.1 au, separated in time by $dt_{enc}=$ 1,000, 10,000 and 100,000 years (see section \ref{sect:follow-on}).

\begin{figure*}
    \centering
    \includegraphics[width=.45\linewidth]{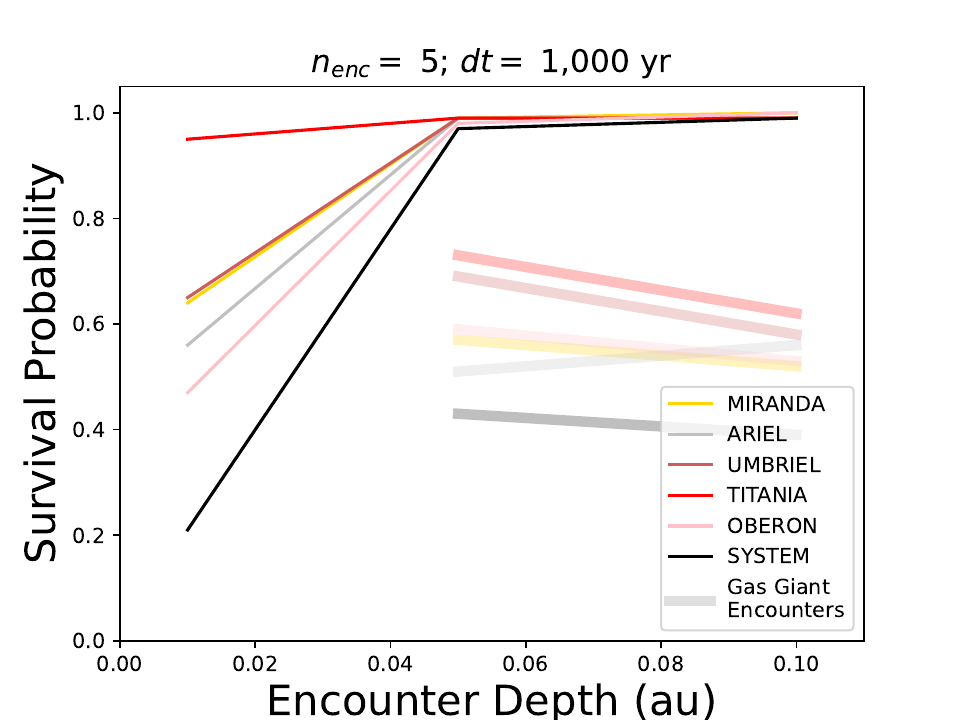}
    \includegraphics[width=.45\linewidth]{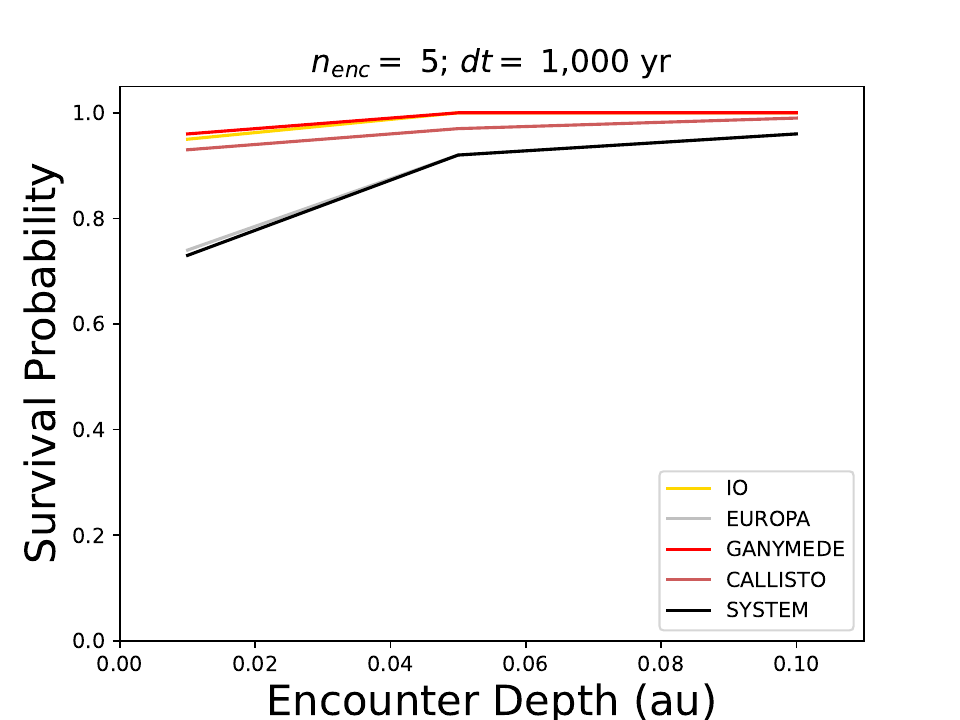}
    \\
    \includegraphics[width=.45\linewidth]{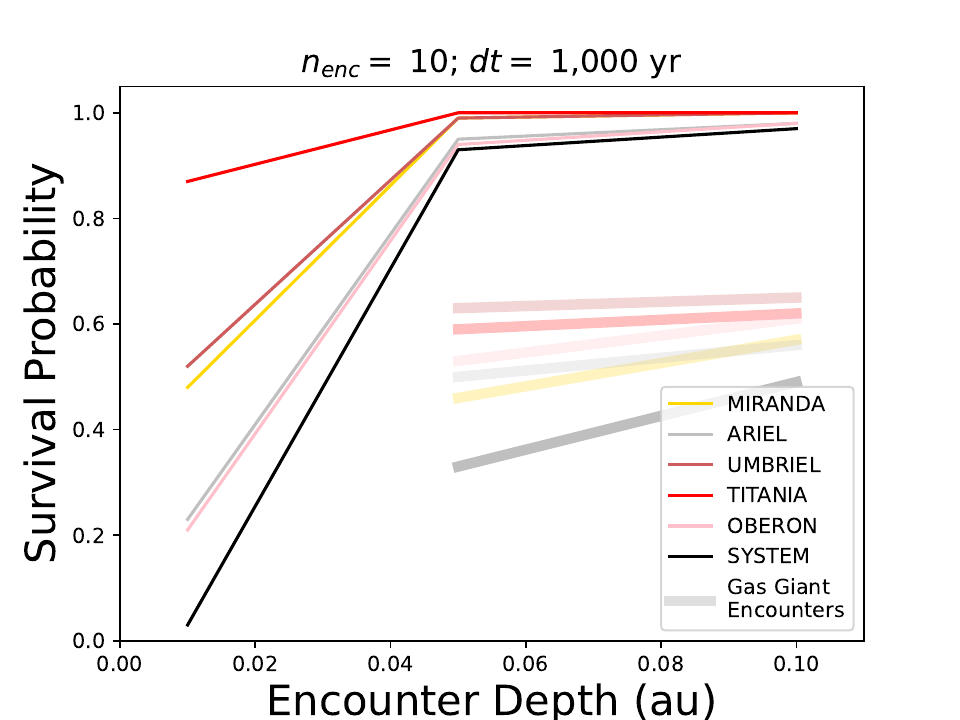}
    \includegraphics[width=.45\linewidth]{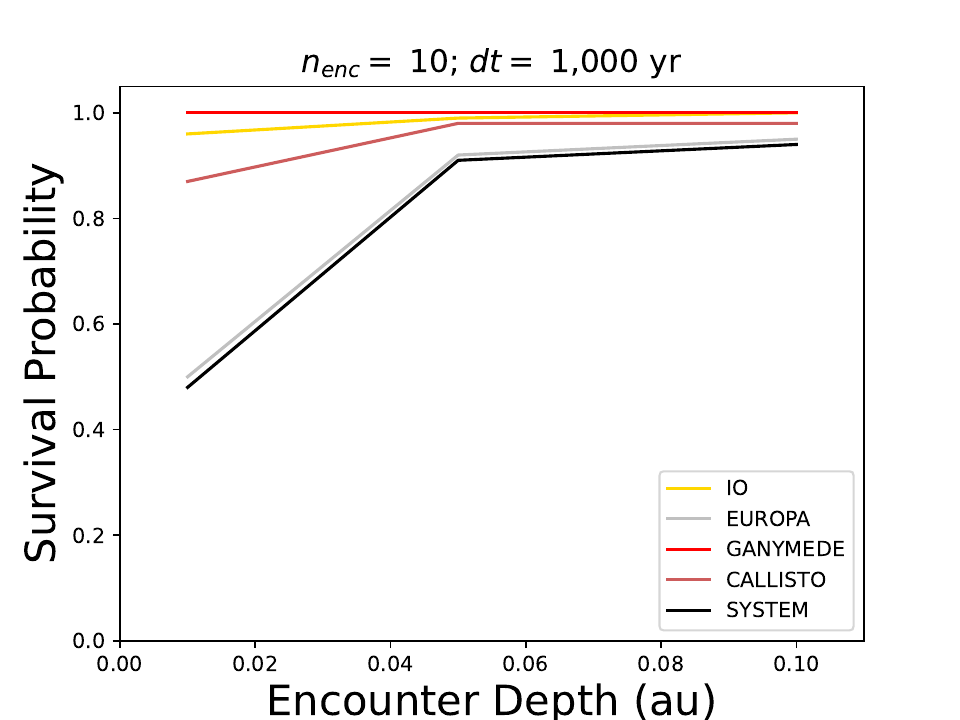}
    \\
    \includegraphics[width=.45\linewidth]{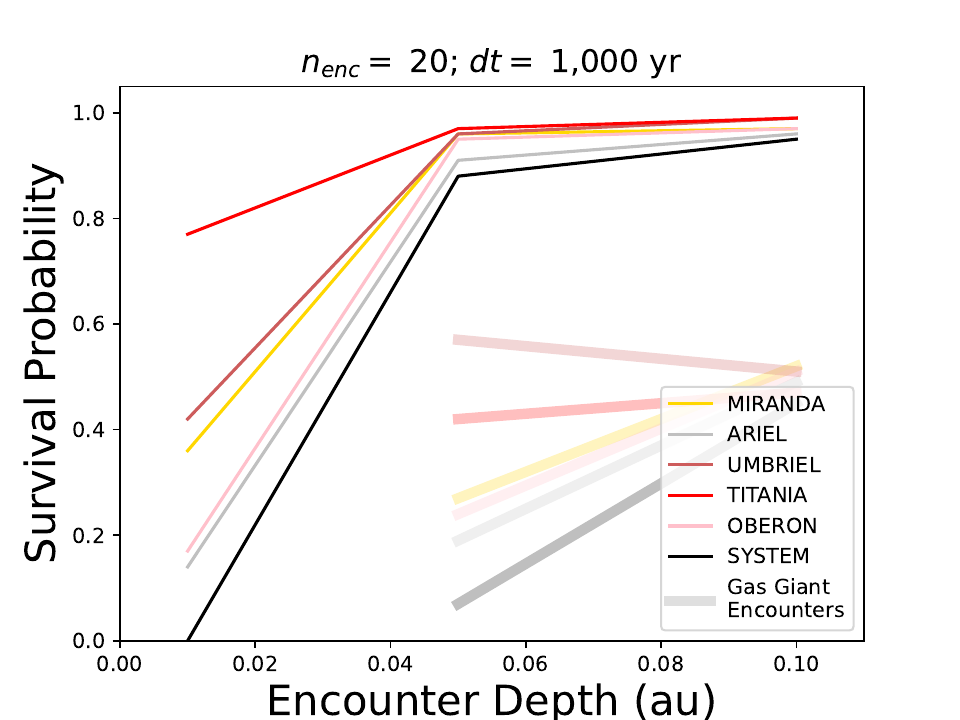}
    \includegraphics[width=.45\linewidth]{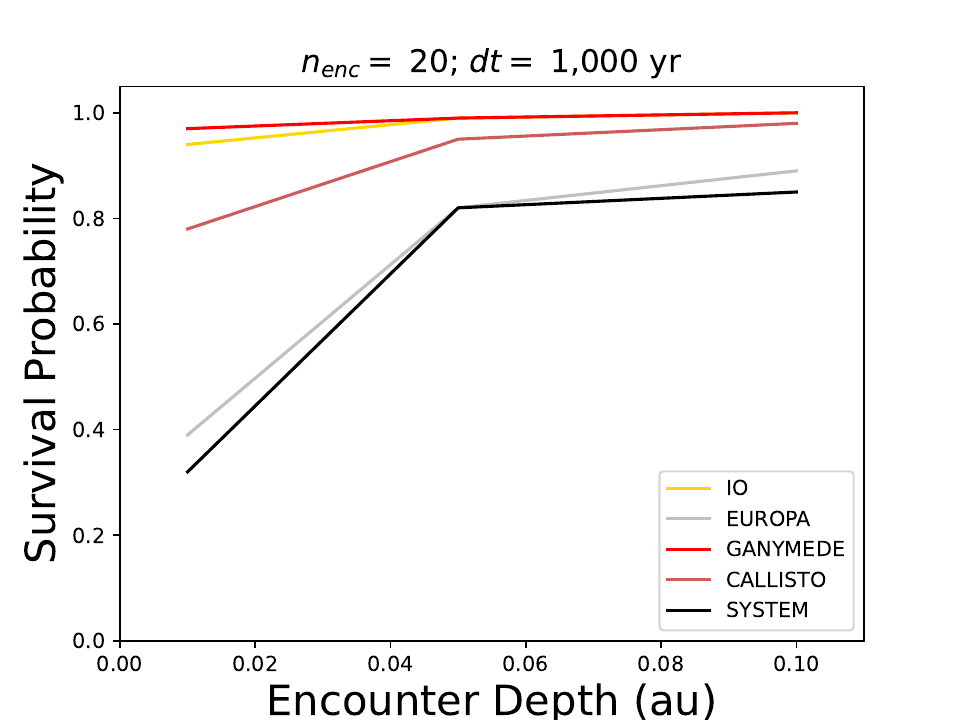}
    \caption{Fraction of systems that were stable in a set of our simulations investigating the effects of repeated encounters at various depths with ice giants (thin lines) and gas giants (thick lines).  The left panels show data for the Uranian system, and the right panels depict results of simulations focused on the Jovian system.  All systems were perturbed by $n_{enc}$ close encounters (5, 10 and 20 for the upper, middle and lower panels, respectively) separated by 1,000 years in time.}
    \label{fig:follow_on}
\end{figure*}

Figure \ref{fig:follow_on} plots the results of our simulations testing $dt_{enc}=$ 1,000 years.  The different panels show different values of $N_{enc}$ for simulations of Uranus' (left panels) and Jupiter's moons (right panels).  The different lines give the survival probability of different moons for different average encounter distances, $D_{enc}$.  Since the vast majority of moons are removed via collision with a larger moon, the differences between the lines reflect both the strength of the instability that takes place within the satellite system, and the sensitivity of each particular moon to the disturbance.  It is unsurprising that Europa -- the least massive of the Galilean satellites -- is most frequently lost since it is always the smallest body in a collision and is thus merged with the larger body\footnote{We remind the reader that we simply merge two bodies when they collide, keeping the more massive of the two alive.  In reality, many of these collisions would likely result in the disruption of both moons, see section \ref{sect:consequences}}.  It is interesting that Io is more likely to survive than Callisto.  This is because the two most common types of collision sequences around Jupiter in our simulation suite are (i) Io collides with Europa while Callisto and Ganymede survive on excited orbits and (ii) Io collides with Europa and Callisto collides with Ganymede.  Similarly, at Uranus we find that Umbriel-Ariel collisions are actually more common than Ariel-Miranda mergers, and Oberon tends to be the first satellite to be lost via collision with Titania.

It is clear from figure \ref{fig:follow_on} that, for all three tested encounter depths, $D_{enc}$, the total number of successive encounters can significantly affect system survival.  For ice giant encounters with Uranus, we find that unstable evolutionary tracks occur only 1\% of the time after 5 consecutive encounters at $D_{enc}=$ 0.1 au.  This fraction increases to 3\% after 10 encounters, and 5\% after 20 encounters.  For a series of encounters at 0.01 au, the effects of repeated encounters are even more drastic.  79\% of our simulations destabilized after 5 such encounters, and this increased to 97\% for 10 encounters and 100\% for 20.  While each evolution is necessarily unique, and it would be impractical to study the details and intricacies of why each particular sequence either did or did not lead to collisions between the moons, broadly speaking the primary factor that is predictive of system stability is the degree to which the eccentricity of the outer moons is excited by the close encounters.  At Uranus, for example, if Oberon and Titania's eccentricities are each excited to greater than around 0.08-0.1 as a result of any of the encounters, the excitation quickly bleeds throughout the system, greatly enhancing secular forcing on the smaller moons and quickly leading to collisions within a few kyr. 

\begin{figure}
    \centering
    \includegraphics[width=0.99\linewidth]{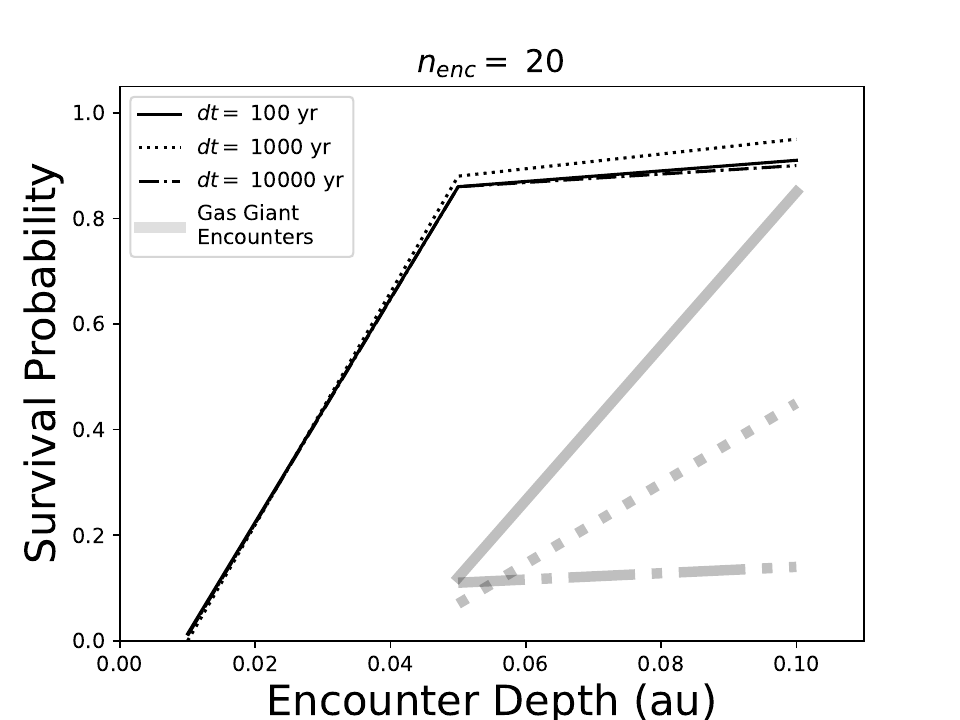}
    \caption{Same as figure \ref{fig:follow_on}, but showing the effects of different time spacings between encounters (dt) on the survival of the Uranian moons.}
    \label{fig:dt_enc}
\end{figure}

For the majority of $N_{enc}$/$D_{enc}$ combinations, varying $dt_{enc}$ does not meaningfully affect the probability of satellite survival.  However, we do note several instances where the smaller values of $dt_{enc}=$ 100 and 1,000 years yield more stable systems than the largest value (10,000 years).  This is most clearly the case for deep, $D_{enc}=$ 0.01 and 0.05 au ice giant encounters with Uranus or Jupiter.  For instance, our simulations that subjected Jupiter's moons to 20 encounters at 0.01 au had a survival probability of 42\% for $dt_{enc}=$ 100 yr, 32\% for $dt_{enc}=$ 1,000 yr, and 27\% for $dt_{enc}=$ 10,000 yr.  While 20 encounters at 0.01 au always destabilizes Uranus's moons, five encounters is not enough to do so, and we see that this same trend manifests in the results.  88\% of these systems are stable for $dt_{enc}=$ 100 yr, compared to 79\% for $dt_{enc}=$ 1,000 yr, and 75\% for $dt_{enc}=$ 10,000 yr.  This trend also holds for chains of encounters between Uranus and the gas giants as well.  This can clearly be seen in figure \ref{fig:dt_enc}, which plots the Uranian system survival probability for all of our $N_{enc}=$ 20 simulations in a manner similar to figure \ref{fig:follow_on}. 

We attribute this minor trend of increased stability for lower $dt_{enc}$ to the fact that the full encounter duration occurs on a timescale of order that of the secular timescale.  Indeed, the secular eigenfrequencies and the apsidal/nodal precession rates in the modern Uranian system are of order $\sim$0.2-20 $^{\circ}$/$yr$ \citep{malhotra89,cuk20}.  Of particular relevance to the typical destabilization mechanism in our simulations -- excitation of Oberon's eccentricity during the encounter -- the current period of the fifth eigenfrequency in the system (the one associated with Oberon's eccentricity), $g_{5}$ is 972 years.  Thus, if the entire epoch of deep encounters ensues in less than a few thousand years, depending on the dynamics of the encounter and how it immediately perturbs each moon's orbit, there is often not enough time for the dynamical effects of one encounter to fully spread through the system before the next encounter ensues.

In conclusion, our additional simulations described in this section, along with those of \ref{sect:ur_results} demonstrate that, while the depth of the deepest planetary encounter is the strongest predictor of satellite system survival, the total number of successive encounters also plays an extremely important role.  In contrast, the time between successive encounters matters slightly for some cases, but our data set is insufficient to conclude that this is any more than a second or third order effect.  For the Uranian moons, we find that, while $\lesssim$5 encounters with other ice giants at $D_{enc}=$ 0.01 au are likely to destroy the system, 10 such encounters are required to essentially guarantee collisions occur between the moons.  This is significant because chains of 10 or more deep planetary encounters at any of the outer planets are actually common in simulations of the Nice Model instability, and we find similar chains in around 30$\%$ of our sample of Uranian encounter histories (e.g. figure \ref{fig:nenc_stats}).  In contrast, we find that reasonable rates of Jovian system integrity across all parameters tested.  Thus, multiple encounters deeper than 0.01 au would be required to achieve a 0\% survival probability.  Since deep encounters at Jupiter are still very common in our sample of instabilities (figure \ref{fig:denc_stats}), this effect does not manifest particularly clearly in our bulk survival rates. 

\subsection{Consequences of Destabilization}
\label{sect:consequences}

In future work, we plan to study the moon-on-moon collisions themselves in greater detail, using hydrodynamical impact models \citep[e.g.][]{wadsley04,reufer12,schaller24}.  Nevertheless, it is worth briefly discussing what the consequences of the collisions produced in this work are likely to be.  In particular, we wish to stress that our results should not be interpreted as evidence that the Nice Model instability could not have taken place in the solar system because it would have destroyed the Uranian and Jovian satellite systems.  Not only do we find cases where the moons are unaffected, but it is also entirely possible that the pre-instability satellite systems of each giant planet did not look the way they do today, and that they were altered via tidal evolution,  satellite mergers or mass transfer during hit-and-run collisions that occurred as a consequence of planetary encounters during the instability.  Indeed, even if the orbits of each primordial satellite were excited to the point of collision with another major satellite during the instability, it is difficult to envision a scenario where a significant amount of satellite mass is lost from the system via ejection or merger with the planet.  Even if a sizable fraction of a satellite's mass was ground to dust in the aftermath of a high-velocity impact, it is likely that the ejected mass would reaccrete relatively rapidly due to the short collisional timescales in giant planet satellite systems \citep[e.g.][]{mosqueira03a,mosqueira03b}.

The majority of the collisions in our simulations occur at geometries that would actually result in a hit-and-run \citep[e.g.][]{genda12}.  Collisions that might lead to catastrophic disruption of the colliding bodies are fairly rare, but still present in our collection of simulations.   The most violent event in our sample -- in terms of being head-on at high velocity -- was a 1.61 km/s -- over 13 times the two body escape velocity -- collision between Titania and Oberon at an impact angle of 14.3$^{\circ}$ (computed from the velocity vectors output at the time of collision).  The majority of head-on collisions, however, occur at less extreme velocities.  Specifically, those involving the smaller inner moons Miranda, Ariel and Umbriel typically occur at 1-2 km/s, where as the outer moons tend to collide at around 0.5-1 km/s.  This is comparable to the threshold for significant impact vaporization via shock compression for collisions between icy planetesimals \protect\citep[$\sim$1 km/s:][]{stewart08,davies19}.  Therefore, the most likely consequence of Uranian satellite destabilization seems to be a series of hit-and-run impacts that result in volatile vaporization and some potential fragmentation, rather than satellite consolidation.  If the colliding bodies hit-and-run at a velocity that does not cause significant disruption, they would likely experience additional follow-on collisions -- vaporizing additional ice in the process -- up until the point that they collide head-on and totally disrupt, or their orbits tidally damp to the point that they no longer cross.  If the Uranian satellites lost a significant fraction of ice through this process, an explanation as to why they are icier than Saturn's satellites would be required, although our results certainly indicate that a similar instability in the Saturnian system might have transpired as well.  In any case, much of the disrupted material would eventually re-accrete onto the surviving objects; though it is certainly easy to envision collisional scenarios where material is exchanged between moons, or preferentially eroded.  While this might provide an interesting potential explanation for Miranda's composition, we leave the exploration of that possibility to future work.

It is also possible that other qualities of the current moon systems might trace their origins to a post-formation epoch of encounters and collisions.  For instance, if the primordial system contained equal-mass satellites, a more heterogeneous mass distribution might result if certain moons were disrupted by more energetic collisions than others.  Similarly, mass loss to the central planet would presumably be more efficient in the inner parts of the satellite system, thus promoting the re-accreted system to possess a hierarchical mass distribution akin to the current Uranian system.

\subsection{Comparison with Previous Works}

As noted in the introduction, the two most relevant points of comparison for our current study are \citet{deienno11} and \citet{deienno14}.  Here, we find a much smaller fraction of the tested parameter space allows for satellite system survival.  However, given the differences in methodology and scope, we still conclude that our results are largely consistent with the previously published ones.  \citet{deienno11} considered an older version of the Nice Model \citet{gomes05} with just four initial planets.  In these evolutionary tracks, Uranus occasionally experiences a small number of close encounters with Neptune, however most sequences avoid such events entirely.  Thus, the primary difference between that work and our current study is the assumed frequency of planetary encounters, which we take to be high in accordance with the current consensus version of the Nice Model \citep[which requires the encounters for a number of reasons, most importantly to aid in the capture of irregular satellites:][]{nesvorny14a}.

\citet{deienno14} tested three possible instability evolutionary tracks at Jupiter, far less than the 122 considered in this work.  The most successful case did not contain any planetary encounters with $D_{enc}<$ 0.05 au, and the Galilean satellites' orbits were not appreciably affected in any of the trials.  Here, we also find cases of quiescent sequences where Jupiter avoids deep encounters, the final orbits of the giant planets still reasonably match the real ones (section \ref{sect:methods1}), and the moon system and Laplace resonance remain intact after the instability.  26\% of our sample of instabilities do not contain encounters between Jupiter and other planets with $D_{enc}<$ 0.05 au (see figure \ref{fig:nenc_stats}).  While this is consistent with \citet{deienno14} testing one of three such cases, we find that only around half of these less violent evolutionary tracks actually result in the stability of the Jovian moons.  The reason for this is the effect of repeated encounters with 0.05 $\lesssim D_{enc} \lesssim$ 0.1 au (figure \ref{fig:follow_on}), and the successful instability tested in \citet{deienno14} only contained 3 such encounters.  Thus, we conclude that our results are also consistent with the study of \citet{deienno14}, and we expand on those results by demonstrating a rather appreciable degree of mutual exclusivity between the sample of instabilities that allow for Jupiter's moons to survive, and those that don't disrupt Uranus'. 

\section{Discussion}

Our paper largely focuses on the consequences of the giant planet instability at Uranus, rather than Jupiter or Saturn.  In the case of Jupiter, this is because the modern presence of the Laplace resonance does not make realizations where the primordial Galilean satellites are significantly perturbed particularly relevant to the solar system.  On the other hand, a scenario where the Saturnian satellites are destabilized early in the solar system's history is not precluded by such a constraint, and we plan to develop this scenario further in a forthcoming manuscript.  Moreover, while the Jovian and Uranian total encounter numbers and the minimum encounter distances were anti-correlated in our 5GP and 6GP samples, this trend was not as clear in the statistics for Uranus and Saturn.  Thus, we chose to present the Jovian case here to more clearly demonstrate how the types of instabilities that tend to promote satellite survival at Uranus are often the worst cases for Jupiter and the Laplace resonance.

It is nevertheless tantalizing to envision a scenario where a particular series of planetary encounters during the instability is responsible for dislodging Callisto from a natal 2:1 resonance with Ganymede, exciting the inclination of Iapetus \citep{nesvorny14b}, breaking up or consolidating other primordial moons of Saturn, capturing Triton and disrupting any other primordial satellites of Neptune.  However, if our sample of instabilities is any indication, identifying such an ideal evolution would likely be quite challenging.

If planetary encounters indeed occurred after nebular gas dissipation, the only viable evolutionary tracks in our sample are those with very few moderately deep encounters at Jupiter \citep{deienno14}.  The survival of the primordial satellite systems of the other three planets is then far less stringent of a constraint, and our results strongly suggest that the current version of the Nice Model is highly likely to result in primordial satellite system destabilization at one or more of the other giant planets.  We do not view this as particularly problematic since multiple models invoke dynamical sculpting of these systems for one reason or another after gas disk dissipation \citep[e.g.][]{nesvorny07,morby12_uranus,nesvorny14b}.  

We have shown here that the types of instabilities that best maximize the probability of Jupiter avoiding deep encounters during the instability also tend to subject Uranus to a rather violent and lengthy epoch of encounters.  These results have three potential implications: (1) the Uranian satellites were destabilized to the point of collisions at least twice (once as a result of the tilting impact and once more during the giant planet instability), (2) the current version of the Nice Model must be revised \citep[see also][]{hunt22,edwards24,raorane24} or (3) the solar system is the result of fairly unlikely instability evolution that entailed almost no deep encounters between Uranus and the other giant planets.  We tend to favor the first or third explanations.  Scenario (1) presents an interesting mechanism -- high-velocity hit-and-run collisions -- that could be applied in future studies targeting other peculiarities in the system, for instance Miranda's inventories of rock and ice.  Moreover, even if the first scenario eventually proves unworkable \citep[e.g., incompatible with models for tilting Uranus and reaccreting its moons,][]{salmon22,rufu22}, in absence of a mechanism other than planetary encounters being proposed to explain irregular satellite capture, the low-probability interpretation of (3) is more straight forward than completely discarding the Nice Model framework, and also potentially consistent with studies that infer ancient ages for the Moon's surfaces \citep{wong23,bottke24}.


\section{Conclusions}

In this paper we studied the effects of planetary encounters that occurred during the solar system's giant planet instability \citep{tsiganis05,nesvorny12} on the regular satellite systems of Jupiter and Uranus.  Building on previous work on the topic \citep{deienno11,deienno14}, our study greatly increases the sample of potential instability evolutionary tracks evaluated against giant planet satellite system stability constraints.  In this manner, our study includes encounter sequences derived from instabilities utilizing the full spectrum of proposed outer solar system initial conditions \citep{clement21_instb,clement21_instb2}.  The primary conclusions of our work are as follows:

\begin{enumerate}
    \item Deep planetary encounters at each modern giant planet are fairly ubiquitous in our sample.  The probability of experiencing an encounter within 0.02 au with an ice giant-mass planet in any given realization of the instability is $>$ 50\% for all four outer planets.
    \item Uranus' moons are particularly sensitive to the effects of planetary encounters.  Specifically, 91\% of the 122 instability evolutionary tracks we tested destabilized Uranus' moons in at least half of our simulations.  By virtue of its location in the middle of the pre-instability giant planet configuration, it tends to experience encounters that are deeper and more frequent.  Moreover, the system itself is more likely to destabilize following an encounter of a given depth than the Galilean satellite system.
    \item The probability of \textit{both Jupiter and Uranus' moons surviving the same instability} is quite low -- around 1\%.  One reason for this is that Uranus's moons are more likely to survive in instabilities initialized with three total ice giants, and the opposite is true for Jupiter.  In general, such instabilities are brief, well-behaved, and quite similar to the favored types of evolutions in the ''jumping Jupiter'' instability model \citep{nesvorny11} where a single deep encounter with the additional ice giant leads to a brief instability and a step-wise divergence of Jupiter and Saturn's semi-major axes.  In these sequences, Jupiter's moons tend to destabilize precipitously.  However, occasionally Uranus' satellites can ``get lucky'' and avoid particularly deep encounters.  While our results are not sufficient to rule out the five planet version of the Nice Model on account of Jovian satellite survival, it is still noteworthy that the six planet version more consistently produces survivable encounter sequences.  However, the only instances in our study where Jupiter and Uranus simultaneously survived \textit{the same} instability came with five initial planets.
    \item If we accept that we must chose an instability that perseveres Jupiter's satellites -- and thus concede that there is a high likelihood of Uranus' moons being disrupted -- our results suggest that the Uranian system was excited to the point of collisions at least twice after its formation.  Once as a result of the tilting impact \citep{morby12_uranus}, and a second time as a result of the Nice Model instability. The most likely result of latter destabilization is a series of hit-and-run collisions between neighboring satellites.
\end{enumerate}

We conclude by emphasizing the fact that simulations of the Nice Model instability are highly chaotic and stochastic.  Planetary encounters, and the semi-major axis jumps that result from them, are thought to have played a key role in sculpting many small body populations throughout the solar system.  It is highly likely that none of the modeled instabilities in the literature contain the precise sequences of encounters necessary to exactly reproduce all aspects of the solar system.  While it is certainly possible that all four primordial regular satellite systems in the outer solar system were unaffected by planetary encounters, our results strongly suggest that this is not the case.  Thus, our results should serve as motivation for future investigation of the consequences of potential dynamical instabilities ensuing in the giant planet moon systems as a result of the Nice Model instability.

\section*{Acknowledgments}
The authors thank Ramon Brasser and a second anonymous reviewer for thoughtful and insightful reviews that greatly improved the quality of the manuscript.  This research was supported by the Texas Advanced Computing Center (TACC) through the Frontera project.  MSC is supported by NASA Emerging Worlds grants 80NSSC23K0868 and 80NSSC25K0348, as well as NASA’s CHAMPs team, supported by NASA under Grant No. 80NSSC21K0905 issued through the Interdisciplinary Consortia for Astrobiology Research (ICAR) program.  NAK and AI are supported via NASA EW Grant 80NSSC23K0868.  The work of R.D. was partially supported by the NASA Emerging Worlds program, grant 80NSSC25K0348, and by the Center for Lunar Origin and Evolution (CLOE), a team in NASA's SSERVI program (cooperative agreement 80NSSC23M0176).

\bibliographystyle{apj}
\newcommand{\sci}{$Science$ }

\bibliography{newbib.bib}

\begin{thebibliography}{}
\expandafter\ifx\csname natexlab\endcsname\relax\def\natexlab#1{#1}\fi

\bibitem[{{Agnor} \& {Hamilton}(2006)}]{agnor06}
{Agnor}, C.~B., \& {Hamilton}, D.~P. 2006, \nat, 441, 192

\bibitem[{{Batygin} \& {Morbidelli}(2020)}]{batygin_morby20}
{Batygin}, K., \& {Morbidelli}, A. 2020, \apj, 894, 143

\bibitem[{{Bitsch} {et~al.}(2015){Bitsch}, {Johansen}, {Lambrechts}, \& {Morbidelli}}]{bitch15}
{Bitsch}, B., {Johansen}, A., {Lambrechts}, M., \& {Morbidelli}, A. 2015, \aap, 575, A28

\bibitem[{{Bolmont} {et~al.}(2015){Bolmont}, {Raymond}, {Leconte}, {Hersant}, \& {Correia}}]{bolmont15}
{Bolmont}, E., {Raymond}, S.~N., {Leconte}, J., {Hersant}, F., \& {Correia}, A. C.~M. 2015, \aap, 583, A116

\bibitem[{{Boss}(1997)}]{boss97}
{Boss}, A.~P. 1997, Science, 276, 1836

\bibitem[{{Bottke} {et~al.}(2024){Bottke}, {Vokrouhlick{\'y}}, {Nesvorn{\'y}}, {Marschall}, {Morbidelli}, {Deienno}, {Marchi}, {Kirchoff}, {Dones}, \& {Levison}}]{bottke24}
{Bottke}, W.~F., {Vokrouhlick{\'y}}, D., {Nesvorn{\'y}}, D., {et~al.} 2024, \psj, 5, 88

\bibitem[{{Bou{\'e}} \& {Laskar}(2010)}]{boue10}
{Bou{\'e}}, G., \& {Laskar}, J. 2010, \apjl, 712, L44

\bibitem[{{Brasser} {et~al.}(2009){Brasser}, {Morbidelli}, {Gomes}, {Tsiganis}, \& {Levison}}]{brasser09}
{Brasser}, R., {Morbidelli}, A., {Gomes}, R., {Tsiganis}, K., \& {Levison}, H.~F. 2009, \aap, 507, 1053

\bibitem[{Brasser {et~al.}(2020)Brasser, Werner, \& Mojzsis}]{brasser20}
Brasser, R., Werner, S., \& Mojzsis, S. 2020, Icarus, 338, 113514

\bibitem[{{Brown} {et~al.}(1998){Brown}, {Koresko}, \& {Blake}}]{brown98_nereid}
{Brown}, M.~E., {Koresko}, C.~D., \& {Blake}, G.~A. 1998, \apjl, 508, L175

\bibitem[{{Canup} \& {Ward}(2002)}]{canup02}
{Canup}, R.~M., \& {Ward}, W.~R. 2002, \aj, 124, 3404

\bibitem[{{Canup} \& {Ward}(2006)}]{canup06}
---. 2006, \nat, 441, 834

\bibitem[{{Chau} {et~al.}(2021){Chau}, {Reinhardt}, {Izidoro}, {Stadel}, \& {Helled}}]{chau21}
{Chau}, A., {Reinhardt}, C., {Izidoro}, A., {Stadel}, J., \& {Helled}, R. 2021, \mnras, 502, 1647

\bibitem[{{Clement} {et~al.}(2023){Clement}, {Chambers}, {Kaib}, {Raymond}, \& {Jackson}}]{clement23_merc5}
{Clement}, M.~S., {Chambers}, J.~E., {Kaib}, N.~A., {Raymond}, S.~N., \& {Jackson}, A.~P. 2023, \icarus, 394, 115445

\bibitem[{{Clement} {et~al.}(2021{\natexlab{a}}){Clement}, {Deienno}, {Kaib}, {Izidoro}, {Raymond}, \& {Chambers}}]{clement21_instb2}
{Clement}, M.~S., {Deienno}, R., {Kaib}, N.~A., {et~al.} 2021{\natexlab{a}}, \icarus, 367, 114556

\bibitem[{{Clement} {et~al.}(2018){Clement}, {Kaib}, {Raymond}, \& {Walsh}}]{clement18}
{Clement}, M.~S., {Kaib}, N.~A., {Raymond}, S.~N., \& {Walsh}, K.~J. 2018, \icarus, 311, 340

\bibitem[{{Clement} {et~al.}(2021{\natexlab{b}}){Clement}, {Raymond}, {Kaib}, {Deienno}, {Chambers}, \& {Izidoro}}]{clement21_instb}
{Clement}, M.~S., {Raymond}, S.~N., {Kaib}, N.~A., {et~al.} 2021{\natexlab{b}}, \icarus, 355, 114122

\bibitem[{{Colombo} \& {Franklin}(1971)}]{columbo71}
{Colombo}, G., \& {Franklin}, F.~A. 1971, \icarus, 15, 186

\bibitem[{{{\'C}uk} {et~al.}(2016){{\'C}uk}, {Dones}, \& {Nesvorn{\'y}}}]{cuk16}
{{\'C}uk}, M., {Dones}, L., \& {Nesvorn{\'y}}, D. 2016, \apj, 820, 97

\bibitem[{{{\'C}uk} {et~al.}(2026){{\'C}uk}, {El Moutamid}, {Fuller}, \& {Lainey}}]{cuk26}
{{\'C}uk}, M., {El Moutamid}, M., {Fuller}, J., \& {Lainey}, V. 2026, \psj, 7, 52

\bibitem[{{{\'C}uk} {et~al.}(2020){{\'C}uk}, {El Moutamid}, \& {Tiscareno}}]{cuk20}
{{\'C}uk}, M., {El Moutamid}, M., \& {Tiscareno}, M.~S. 2020, \psj, 1, 22

\bibitem[{{Cuzzi} \& {Estrada}(1998)}]{cuzziandestrada98}
{Cuzzi}, J.~N., \& {Estrada}, P.~R. 1998, \icarus, 132, 1

\bibitem[{{Davies} {et~al.}(2019){Davies}, {Root}, {Carter}, {Duncan}, {Spaulding}, {Kraus}, {Stewart}, \& {Jacobsen}}]{davies19}
{Davies}, E.~J., {Root}, S., {Carter}, P.~J., {et~al.} 2019, in 50th Annual Lunar and Planetary Science Conference, Lunar and Planetary Science Conference, 1257

\bibitem[{{de Kleer} {et~al.}(2024){de Kleer}, {Hughes}, {Nimmo}, {Eiler}, {Hofmann}, {Luszcz-Cook}, \& {Mandt}}]{dekleer24}
{de Kleer}, K., {Hughes}, E.~C., {Nimmo}, F., {et~al.} 2024, Science, 384, 682

\bibitem[{{Deienno} {et~al.}(2018){Deienno}, {Izidoro}, {Morbidelli}, {Gomes}, {Nesvorn{\'y}}, \& {Raymond}}]{deienno18}
{Deienno}, R., {Izidoro}, A., {Morbidelli}, A., {et~al.} 2018, \apj, 864, 50

\bibitem[{{Deienno} {et~al.}(2017){Deienno}, {Morbidelli}, {Gomes}, \& {Nesvorn{\'y}}}]{deienno17}
{Deienno}, R., {Morbidelli}, A., {Gomes}, R.~S., \& {Nesvorn{\'y}}, D. 2017, \aj, 153, 153

\bibitem[{{Deienno} {et~al.}(2024){Deienno}, {Nesvorn{\'y}}, {Clement}, {Bottke}, {Izidoro}, \& {Walsh}}]{deienno24}
{Deienno}, R., {Nesvorn{\'y}}, D., {Clement}, M.~S., {et~al.} 2024, \psj, 5, 110

\bibitem[{{Deienno} {et~al.}(2014){Deienno}, {Nesvorn{\'y}}, {Vokrouhlick{\'y}}, \& {Yokoyama}}]{deienno14}
{Deienno}, R., {Nesvorn{\'y}}, D., {Vokrouhlick{\'y}}, D., \& {Yokoyama}, T. 2014, \aj, 148, 25

\bibitem[{{Deienno} {et~al.}(2011){Deienno}, {Yokoyama}, {Nogueira}, {Callegari}, \& {Santos}}]{deienno11}
{Deienno}, R., {Yokoyama}, T., {Nogueira}, E.~C., {Callegari}, N., \& {Santos}, M.~T. 2011, \aap, 536, A57

\bibitem[{{Duncan} {et~al.}(1998){Duncan}, {Levison}, \& {Lee}}]{duncan98}
{Duncan}, M.~J., {Levison}, H.~F., \& {Lee}, M.~H. 1998, \aj, 116, 2067

\bibitem[{{Duncan} \& {Lissauer}(1997)}]{duncanandlissauer97}
{Duncan}, M.~J., \& {Lissauer}, J.~J. 1997, \icarus, 125, 1

\bibitem[{{Edwards} {et~al.}(2024){Edwards}, {Keller}, {Newton}, \& {Stewart}}]{edwards24}
{Edwards}, G.~H., {Keller}, C.~B., {Newton}, E.~R., \& {Stewart}, C.~W. 2024, Nature Astronomy, 8, 1264

\bibitem[{{Ellithorpe} \& {Kaib}(2022)}]{ellithorpe22}
{Ellithorpe}, E.~A., \& {Kaib}, N.~A. 2022, \mnras, 515, 2914

\bibitem[{{Fernandez} \& {Ip}(1984)}]{fernandez84}
{Fernandez}, J.~A., \& {Ip}, W.-H. 1984, \icarus, 58, 109

\bibitem[{{Fuller} {et~al.}(2016){Fuller}, {Luan}, \& {Quataert}}]{fuller16}
{Fuller}, J., {Luan}, J., \& {Quataert}, E. 2016, \mnras, 458, 3867

\bibitem[{{Genda} {et~al.}(2012){Genda}, {Kokubo}, \& {Ida}}]{genda12}
{Genda}, H., {Kokubo}, E., \& {Ida}, S. 2012, \apj, 744, 137

\bibitem[{{Goldreich} \& {Soter}(1966)}]{goldreich66}
{Goldreich}, P., \& {Soter}, S. 1966, \icarus, 5, 375

\bibitem[{{Gomes} {et~al.}(2005){Gomes}, {Levison}, {Tsiganis}, \& {Morbidelli}}]{gomes05}
{Gomes}, R., {Levison}, H.~F., {Tsiganis}, K., \& {Morbidelli}, A. 2005, \nat, 435, 466

\bibitem[{{Gomes} \& {Morbidelli}(2024)}]{gomes24}
{Gomes}, R., \& {Morbidelli}, A. 2024, \icarus, 420, 116142

\bibitem[{{Greenberg}(1987)}]{greenberg87}
{Greenberg}, R. 1987, \icarus, 70, 334

\bibitem[{{Haisch} {et~al.}(2001){Haisch}, {Lada}, \& {Lada}}]{haisch01}
{Haisch}, Jr., K.~E., {Lada}, E.~A., \& {Lada}, C.~J. 2001, \apjl, 553, L153

\bibitem[{{Heppenheimer} \& {Porco}(1977)}]{heppenheimer77}
{Heppenheimer}, T.~A., \& {Porco}, C. 1977, \icarus, 30, 385

\bibitem[{{Hunt} {et~al.}(2022){Hunt}, {Theis}, {Rehk{\"a}mper}, {Benedix}, {Andreasen}, \& {Sch{\"o}nb{\"a}chler}}]{hunt22}
{Hunt}, A.~C., {Theis}, K.~J., {Rehk{\"a}mper}, M., {et~al.} 2022, Nature Astronomy, 6, 812

\bibitem[{{Hussmann} {et~al.}(2006){Hussmann}, {Sohl}, \& {Spohn}}]{hussmann06}
{Hussmann}, H., {Sohl}, F., \& {Spohn}, T. 2006, \icarus, 185, 258

\bibitem[{{Iess} {et~al.}(2019){Iess}, {Militzer}, {Kaspi}, {Nicholson}, {Durante}, {Racioppa}, {Anabtawi}, {Galanti}, {Hubbard}, {Mariani}, {Tortora}, {Wahl}, \& {Zannoni}}]{iess19}
{Iess}, L., {Militzer}, B., {Kaspi}, Y., {et~al.} 2019, Science, 364, aat2965

\bibitem[{{Izidoro} {et~al.}(2015){Izidoro}, {Morbidelli}, {Raymond}, {Hersant}, \& {Pierens}}]{izidoro15_ig}
{Izidoro}, A., {Morbidelli}, A., {Raymond}, S.~N., {Hersant}, F., \& {Pierens}, A. 2015, \aap, 582, A99

\bibitem[{{Izidoro} {et~al.}(2025){Izidoro}, {Raymond}, {Kaib}, {Morbidelli}, \& {Isella}}]{izidoro25_capture}
{Izidoro}, A., {Raymond}, S.~N., {Kaib}, N.~A., {Morbidelli}, A., \& {Isella}, A. 2025, Nature Astronomy, 9, 982

\bibitem[{{Jacobson} \& {Park}(2025)}]{jacobson25}
{Jacobson}, R.~A., \& {Park}, R.~S. 2025, \aj, 169, 65

\bibitem[{{Jakub{\'\i}k} {et~al.}(2012){Jakub{\'\i}k}, {Morbidelli}, {Neslu{\v{s}}an}, \& {Brasser}}]{jakubik12}
{Jakub{\'\i}k}, M., {Morbidelli}, A., {Neslu{\v{s}}an}, L., \& {Brasser}, R. 2012, \aap, 540, A71

\bibitem[{{Kaib} {et~al.}(2024){Kaib}, {Parsells}, {Grimm}, {Quarles}, \& {Clement}}]{kaib24}
{Kaib}, N.~A., {Parsells}, A., {Grimm}, S., {Quarles}, B., \& {Clement}, M.~S. 2024, \icarus, 415, 116057

\bibitem[{{Kaib} \& {Raymond}(2024)}]{kaibandraymond24}
{Kaib}, N.~A., \& {Raymond}, S.~N. 2024, \apjl, 962, L28

\bibitem[{{Kaib} \& {Sheppard}(2016)}]{kaib16}
{Kaib}, N.~A., \& {Sheppard}, S.~S. 2016, \aj, 152, 133

\bibitem[{{Kaib} {et~al.}(2018){Kaib}, {White}, \& {Izidoro}}]{kaib18}
{Kaib}, N.~A., {White}, E.~B., \& {Izidoro}, A. 2018, \mnras, 473, 470

\bibitem[{{Kuskov} \& {Kronrod}(2001)}]{kuskov01}
{Kuskov}, O.~L., \& {Kronrod}, V.~A. 2001, \icarus, 151, 204

\bibitem[{{Lainey} {et~al.}(2012){Lainey}, {Karatekin}, {Desmars}, {Charnoz}, {Arlot}, {Emelyanov}, {Le Poncin-Lafitte}, {Mathis}, {Remus}, {Tobie}, \& {Zahn}}]{lainey12}
{Lainey}, V., {Karatekin}, {\"O}., {Desmars}, J., {et~al.} 2012, \apj, 752, 14

\bibitem[{{Lainey} {et~al.}(2020){Lainey}, {Casajus}, {Fuller}, {Zannoni}, {Tortora}, {Cooper}, {Murray}, {Modenini}, {Park}, {Robert}, \& {Zhang}}]{lainey20}
{Lainey}, V., {Casajus}, L.~G., {Fuller}, J., {et~al.} 2020, Nature Astronomy, 4, 1053

\bibitem[{{Lambrechts} {et~al.}(2014){Lambrechts}, {Johansen}, \& {Morbidelli}}]{lambrechts14a}
{Lambrechts}, M., {Johansen}, A., \& {Morbidelli}, A. 2014, \aap, 572, A35

\bibitem[{{Lambrechts} \& {Lega}(2017)}]{lambrechts17}
{Lambrechts}, M., \& {Lega}, E. 2017, \aap, 606, A146

\bibitem[{{Levison} {et~al.}(2015){Levison}, {Kretke}, \& {Duncan}}]{levison15_gp}
{Levison}, H.~F., {Kretke}, K.~A., \& {Duncan}, M.~J. 2015, \nat, 524, 322

\bibitem[{{Lu} \& {Laughlin}(2022)}]{lu22}
{Lu}, T., \& {Laughlin}, G. 2022, \psj, 3, 221

\bibitem[{{Lunine} \& {Stevenson}(1982)}]{lunine82}
{Lunine}, J.~I., \& {Stevenson}, D.~J. 1982, \icarus, 52, 14

\bibitem[{{Madeira} {et~al.}(2021){Madeira}, {Izidoro}, \& {Giuliatti Winter}}]{madeira21}
{Madeira}, G., {Izidoro}, A., \& {Giuliatti Winter}, S.~M. 2021, \mnras, 504, 1854

\bibitem[{{Malhotra}(1991)}]{malhotra91}
{Malhotra}, R. 1991, \icarus, 94, 399

\bibitem[{{Malhotra} {et~al.}(1989){Malhotra}, {Fox}, {Murray}, \& {Nicholson}}]{malhotra89}
{Malhotra}, R., {Fox}, K., {Murray}, C.~D., \& {Nicholson}, P.~D. 1989, \aap, 221, 348

\bibitem[{{Mayer} {et~al.}(2002){Mayer}, {Quinn}, {Wadsley}, \& {Stadel}}]{mayer02}
{Mayer}, L., {Quinn}, T., {Wadsley}, J., \& {Stadel}, J. 2002, Science, 298, 1756

\bibitem[{{Mignard}(1979)}]{mignard79}
{Mignard}, F. 1979, Moon and Planets, 20, 301

\bibitem[{{Morbidelli} {et~al.}(2005){Morbidelli}, {Levison}, {Tsiganis}, \& {Gomes}}]{morby05}
{Morbidelli}, A., {Levison}, H.~F., {Tsiganis}, K., \& {Gomes}, R. 2005, \nat, 435, 462

\bibitem[{{Morbidelli} \& {Nesvorn{\'y}}(2020)}]{morby20_review}
{Morbidelli}, A., \& {Nesvorn{\'y}}, D. 2020, in The Trans-Neptunian Solar System, ed. D.~{Prialnik}, M.~A. {Barucci}, \& L.~{Young}, 25--59

\bibitem[{{Morbidelli} {et~al.}(2018){Morbidelli}, {Nesvorny}, {Laurenz}, {Marchi}, {Rubie}, {Elkins-Tanton}, {Wieczorek}, \& {Jacobson}}]{morby18}
{Morbidelli}, A., {Nesvorny}, D., {Laurenz}, V., {et~al.} 2018, \icarus, 305, 262

\bibitem[{{Morbidelli} {et~al.}(2012){Morbidelli}, {Tsiganis}, {Batygin}, {Crida}, \& {Gomes}}]{morby12_uranus}
{Morbidelli}, A., {Tsiganis}, K., {Batygin}, K., {Crida}, A., \& {Gomes}, R. 2012, \icarus, 219, 737

\bibitem[{{Mosqueira} \& {Estrada}(2003{\natexlab{a}})}]{mosqueira03a}
{Mosqueira}, I., \& {Estrada}, P.~R. 2003{\natexlab{a}}, \icarus, 163, 198

\bibitem[{{Mosqueira} \& {Estrada}(2003{\natexlab{b}})}]{mosqueira03b}
---. 2003{\natexlab{b}}, \icarus, 163, 232

\bibitem[{{Murray} \& {Dermott}(1999)}]{dermott99}
{Murray}, C.~D., \& {Dermott}, S.~F. 1999, {Solar system dynamics}

\bibitem[{{Ndugu} {et~al.}(2018){Ndugu}, {Bitsch}, \& {Jurua}}]{ndugu18}
{Ndugu}, N., {Bitsch}, B., \& {Jurua}, E. 2018, \mnras, 474, 886

\bibitem[{{Nesvorn{\'y}}(2011)}]{nesvorny11}
{Nesvorn{\'y}}, D. 2011, \apjl, 742, L22

\bibitem[{{Nesvorn{\'y}}(2015{\natexlab{a}})}]{nesvorny15a}
---. 2015{\natexlab{a}}, \aj, 150, 73

\bibitem[{{Nesvorn{\'y}}(2015{\natexlab{b}})}]{nesvorny15b}
---. 2015{\natexlab{b}}, \aj, 150, 68

\bibitem[{{Nesvorn{\'y}} \& {Morbidelli}(2012)}]{nesvorny12}
{Nesvorn{\'y}}, D., \& {Morbidelli}, A. 2012, \aj, 144, 117

\bibitem[{{Nesvorn{\'y}} {et~al.}(2021){Nesvorn{\'y}}, {Roig}, \& {Deienno}}]{nesvorny21_tp}
{Nesvorn{\'y}}, D., {Roig}, F.~V., \& {Deienno}, R. 2021, \aj, 161, 50

\bibitem[{{Nesvorn{\'y}} {et~al.}(2023){Nesvorn{\'y}}, {Roig}, {Vokrouhlick{\'y}}, {Bottke}, {Marchi}, {Morbidelli}, \& {Deienno}}]{nesvorny23}
{Nesvorn{\'y}}, D., {Roig}, F.~V., {Vokrouhlick{\'y}}, D., {et~al.} 2023, \icarus, 399, 115545

\bibitem[{{Nesvorn{\'y}} \& {Vokrouhlick{\'y}}(2016)}]{nesvorny16}
{Nesvorn{\'y}}, D., \& {Vokrouhlick{\'y}}, D. 2016, \apj, 825, 94

\bibitem[{{Nesvorny} {et~al.}(2018){Nesvorny}, {Vokrouhlicky}, {Bottke}, \& {Levison}}]{nesvorny18}
{Nesvorny}, D., {Vokrouhlicky}, D., {Bottke}, W.~F., \& {Levison}, H.~F. 2018, ArXiv e-prints, arXiv:1809.04007

\bibitem[{{Nesvorn{\'y}} {et~al.}(2014{\natexlab{a}}){Nesvorn{\'y}}, {Vokrouhlick{\'y}}, \& {Deienno}}]{nesvorny14a}
{Nesvorn{\'y}}, D., {Vokrouhlick{\'y}}, D., \& {Deienno}, R. 2014{\natexlab{a}}, \apj, 784, 22

\bibitem[{{Nesvorn{\'y}} {et~al.}(2014{\natexlab{b}}){Nesvorn{\'y}}, {Vokrouhlick{\'y}}, {Deienno}, \& {Walsh}}]{nesvorny14b}
{Nesvorn{\'y}}, D., {Vokrouhlick{\'y}}, D., {Deienno}, R., \& {Walsh}, K.~J. 2014{\natexlab{b}}, \aj, 148, 52

\bibitem[{{Nesvorn{\'y}} {et~al.}(2007{\natexlab{a}}){Nesvorn{\'y}}, {Vokrouhlick{\'y}}, \& {Morbidelli}}]{nesvory07}
{Nesvorn{\'y}}, D., {Vokrouhlick{\'y}}, D., \& {Morbidelli}, A. 2007{\natexlab{a}}, \aj, 133, 1962

\bibitem[{{Nesvorn{\'y}} {et~al.}(2007{\natexlab{b}}){Nesvorn{\'y}}, {Vokrouhlick{\'y}}, \& {Morbidelli}}]{nesvorny07}
---. 2007{\natexlab{b}}, \aj, 133, 1962

\bibitem[{{Nesvorn{\'y}} {et~al.}(2013){Nesvorn{\'y}}, {Vokrouhlick{\'y}}, \& {Morbidelli}}]{nesvorny13}
---. 2013, \apj, 768, 45

\bibitem[{{Nettelmann} {et~al.}(2013){Nettelmann}, {Helled}, {Fortney}, \& {Redmer}}]{nettelmann13}
{Nettelmann}, N., {Helled}, R., {Fortney}, J.~J., \& {Redmer}, R. 2013, \planss, 77, 143

\bibitem[{{Peale} \& {Lee}(2002)}]{peale02}
{Peale}, S.~J., \& {Lee}, M.~H. 2002, Science, 298, 593

\bibitem[{{Philpott} {et~al.}(2010){Philpott}, {Hamilton}, \& {Agnor}}]{philpott10}
{Philpott}, C.~M., {Hamilton}, D.~P., \& {Agnor}, C.~B. 2010, \icarus, 208, 824

\bibitem[{{Pollack} {et~al.}(1979){Pollack}, {Burns}, \& {Tauber}}]{pollack79}
{Pollack}, J.~B., {Burns}, J.~A., \& {Tauber}, M.~E. 1979, \icarus, 37, 587

\bibitem[{{Pollack} {et~al.}(1996){Pollack}, {Hubickyj}, {Bodenheimer}, {Lissauer}, {Podolak}, \& {Greenzweig}}]{pollack96}
{Pollack}, J.~B., {Hubickyj}, O., {Bodenheimer}, P., {et~al.} 1996, \icarus, 124, 62

\bibitem[{{Porco} {et~al.}(2006){Porco}, {Helfenstein}, {Thomas}, {Ingersoll}, {Wisdom}, {West}, {Neukum}, {Denk}, {Wagner}, {Roatsch}, {Kieffer}, {Turtle}, {McEwen}, {Johnson}, {Rathbun}, {Veverka}, {Wilson}, {Perry}, {Spitale}, {Brahic}, {Burns}, {Del Genio}, {Dones}, {Murray}, \& {Squyres}}]{porco06}
{Porco}, C.~C., {Helfenstein}, P., {Thomas}, P.~C., {et~al.} 2006, Science, 311, 1393

\bibitem[{{Raorane} {et~al.}(2024){Raorane}, {Brasser}, {Matsumura}, {Lau}, {Lee}, \& {Bouvier}}]{raorane24}
{Raorane}, A., {Brasser}, R., {Matsumura}, S., {et~al.} 2024, \icarus, 421, 116231

\bibitem[{{Raymond} {et~al.}(2009){Raymond}, {Armitage}, \& {Gorelick}}]{raymond09_scat}
{Raymond}, S.~N., {Armitage}, P.~J., \& {Gorelick}, N. 2009, \apjl, 699, L88

\bibitem[{{Raymond} {et~al.}(2010){Raymond}, {Armitage}, \& {Gorelick}}]{raymond10}
---. 2010, \apj, 711, 772

\bibitem[{{Reufer} {et~al.}(2012){Reufer}, {Meier}, {Benz}, \& {Wieler}}]{reufer12}
{Reufer}, A., {Meier}, M. M.~M., {Benz}, W., \& {Wieler}, R. 2012, \icarus, 221, 296

\bibitem[{{Rogoszinski} \& {Hamilton}(2020)}]{rogosziniski20}
{Rogoszinski}, Z., \& {Hamilton}, D.~P. 2020, \apj, 888, 60

\bibitem[{{Roig} {et~al.}(2016){Roig}, {Nesvorn{\'y}}, \& {DeSouza}}]{roig16}
{Roig}, F., {Nesvorn{\'y}}, D., \& {DeSouza}, S.~R. 2016, \apjl, 820, L30

\bibitem[{{Rufu} \& {Canup}(2022)}]{rufu22}
{Rufu}, R., \& {Canup}, R.~M. 2022, \apj, 928, 123

\bibitem[{Saillenfest {et~al.}(2022)Saillenfest, Rogoszinski, Lari, Baillié, Boué, Crida, \& Lainey}]{saillenfest22}
Saillenfest, M., Rogoszinski, Z., Lari, G., {et~al.} 2022, Tilting Uranus via the migration of an ancient satellite, doi:10.48550/ARXIV.2209.10590

\bibitem[{{Salmon} \& {Canup}(2022)}]{salmon22}
{Salmon}, J., \& {Canup}, R.~M. 2022, \apj, 924, 6

\bibitem[{{Schaller} {et~al.}(2024){Schaller}, {Borrow}, {Draper}, {Ivkovic}, {McAlpine}, {Vandenbroucke}, {Bah{\'e}}, {Chaikin}, {Chalk}, {Chan}, {Correa}, {van Daalen}, {Elbers}, {Gonnet}, {Hausammann}, {Helly}, {Hu{\v{s}}ko}, {Kegerreis}, {Nobels}, {Ploeckinger}, {Revaz}, {Roper}, {Ruiz-Bonilla}, {Sandnes}, {Uyttenhove}, {Willis}, \& {Xiang}}]{schaller24}
{Schaller}, M., {Borrow}, J., {Draper}, P.~W., {et~al.} 2024, \mnras, 530, 2378

\bibitem[{{Shibaike} {et~al.}(2019){Shibaike}, {Ormel}, {Ida}, {Okuzumi}, \& {Sasaki}}]{shibaike19}
{Shibaike}, Y., {Ormel}, C.~W., {Ida}, S., {Okuzumi}, S., \& {Sasaki}, T. 2019, \apj, 885, 79

\bibitem[{{Slattery} {et~al.}(1992){Slattery}, {Benz}, \& {Cameron}}]{slattery92}
{Slattery}, W.~L., {Benz}, W., \& {Cameron}, A.~G.~W. 1992, \icarus, 99, 167

\bibitem[{{Stewart} {et~al.}(2008){Stewart}, {Seifter}, \& {Obst}}]{stewart08}
{Stewart}, S.~T., {Seifter}, A., \& {Obst}, A.~W. 2008, in 39th Annual Lunar and Planetary Science Conference, Lunar and Planetary Science Conference, 2301

\bibitem[{{Szul{\'a}gyi} {et~al.}(2018){Szul{\'a}gyi}, {Cilibrasi}, \& {Mayer}}]{szulagyi18}
{Szul{\'a}gyi}, J., {Cilibrasi}, M., \& {Mayer}, L. 2018, \apjl, 868, L13

\bibitem[{{Tajeddine} {et~al.}(2017){Tajeddine}, {Nicholson}, {Longaretti}, {El Moutamid}, \& {Burns}}]{tajeddine17}
{Tajeddine}, R., {Nicholson}, P.~D., {Longaretti}, P.-Y., {El Moutamid}, M., \& {Burns}, J.~A. 2017, \apjs, 232, 28

\bibitem[{{Tsiganis} {et~al.}(2005){Tsiganis}, {Gomes}, {Morbidelli}, \& {Levison}}]{tsiganis05}
{Tsiganis}, K., {Gomes}, R., {Morbidelli}, A., \& {Levison}, H.~F. 2005, \nat, 435, 459

\bibitem[{{Wadsley} {et~al.}(2004){Wadsley}, {Stadel}, \& {Quinn}}]{wadsley04}
{Wadsley}, J.~W., {Stadel}, J., \& {Quinn}, T. 2004, \na, 9, 137

\bibitem[{{Wisdom} \& {Holman}(1991)}]{wisdomholman}
{Wisdom}, J., \& {Holman}, M. 1991, \aj, 102, 1528

\bibitem[{{Wong} {et~al.}(2019){Wong}, {Brasser}, \& {Werner}}]{wong19}
{Wong}, E.~W., {Brasser}, R., \& {Werner}, S.~C. 2019, Earth and Planetary Science Letters, 506, 407

\bibitem[{{Wong} {et~al.}(2021){Wong}, {Brasser}, \& {Werner}}]{wong21}
---. 2021, \icarus, 358, 114184

\bibitem[{{Wong} {et~al.}(2023){Wong}, {Brasser}, {Werner}, \& {Kirchoff}}]{wong23}
{Wong}, E.~W., {Brasser}, R., {Werner}, S.~C., \& {Kirchoff}, M.~R. 2023, \icarus, 406, 115763

\bibitem[{{Yap} \& {Batygin}(2025)}]{yap25}
{Yap}, T.~E., \& {Batygin}, K. 2025, \apj, 995, 218

\bibitem[{{Yoder}(1979)}]{yoder79}
{Yoder}, C.~F. 1979, \nat, 279, 767

\bibitem[{{Zahnle} {et~al.}(1998){Zahnle}, {Dones}, \& {Levison}}]{zahnle98}
{Zahnle}, K., {Dones}, L., \& {Levison}, H.~F. 1998, \icarus, 136, 202

\bibitem[{{Zahnle} {et~al.}(2003){Zahnle}, {Schenk}, {Levison}, \& {Dones}}]{zahnle03}
{Zahnle}, K., {Schenk}, P., {Levison}, H., \& {Dones}, L. 2003, \icarus, 163, 263

\bibitem[{{Zellner}(2017)}]{zellner17}
{Zellner}, N.~E.~B. 2017, Origins of Life and Evolution of the Biosphere, arXiv:1704.06694

\end{thebibliography}
\end{document}